\documentclass{article} 
\usepackage{amssymb,amsmath,amsthm} 
\usepackage{epsfig}
\newtheorem{lmm}{Lemma}[section]

\newtheorem{prp}{Proposition}[section]

\newtheorem{thm}{Theorem}[section]

\theoremstyle{definition}

\newtheorem{Exa}{Example}[section]
\newenvironment{exa} {\begin{Exa}}{\qed\end{Exa}}
\newtheorem{Rem}{Remark}[section]
\newenvironment{rem} {\begin{Rem}}{\qed\end{Rem}} 
\newcommand{\etal}{\emph{et al.}}
\newcommand{\ie}{\emph{i.e.}}
\newcommand{\eg}{\emph{e.g.}}
\newcommand{\cf}{\emph{cf}}
\newcommand{\etc}{\emph{etc}}
\newcommand{\demi}{\frac{1}{2}}
\newcommand{\Real}{\mathbb{R}}
\newcommand{\Nat}{\mathbb{N}}
\newcommand{\Int}{\mathbb{Z}}
\newcommand{\Smooth}{C}
\newcommand{\si}{L^1}
\newcommand{\sii}{L^2}
\newcommand{\sinf}{L^\infty}
\newcommand{\sobi}{\mathop{W_0^{1,2}}\nolimits}
\newcommand{\Sobi}{\mathop{W^{1,2}}\nolimits}

\newcommand{\Sobii}{\mathop{W^{2,2}}\nolimits}
\newcommand{\Dom}{\mathop{\mathrm{Dom}}\nolimits}
\newcommand{\supp}{\mathop{\mathrm{supp}}\nolimits}

\newcommand{\Id}{\mathop{\mathit{Id}}\nolimits}
\newcommand{\trans}{I}

\newcommand{\curve}{\Gamma}
\newcommand{\strip}{\Omega}
\newcommand{\stripmap}{\mathcal{L}}
\newcommand{\both}{\iota}
\newcommand{\Hilbert}{\mathcal{H}}
\newcommand{\Hi}{\langle\mathsf{H}\rangle}
\newcommand{\di}{\langle\mathsf{d}\rangle}
\newenvironment{Assumption}[1]
{\begin{description}\item[$#1$]\quad}
{\end{description}}
\title{
\textbf{On the spectrum of curved planar waveguides}
\footnote{To appear in Publ.~RIMS, Kyoto University}
} 
\author{
David \textsc{Krej\v ci\v r\'\i k}$\,^{1}$%
\footnote{ 
Also on leave of absence from
\emph{
Department of Theoretical Physics, Nuclear Physics Institute,
Academy of Sciences, 250\,68 \v{R}e\v{z} near Prague, Czech Republic
}
}
\quad and \quad Jan \textsc{K\v r\'\i \v z}$\,^{2}$
} 
\date{\small
\begin{quote}
\emph{
\begin{itemize}
\item[$^1$]
Departamento de Matem\'atica, Instituto Superior T\'ecnico, \\
Av. Rovisco Pais, 1049-001 Lisboa, Portugal
\item[$^2$]
Department of Physics, Faculty od Education, University of Hradec Kr\'alov\'e, 
Rokitansk\'eho 62, 500 03 Hradec Kr\'alov\'e,
Czech Republic
\item[]
\emph{E-mail:} dkrej@math.ist.utl.pt and jan.kriz@uhk.cz 
\end{itemize}
}
\end{quote}
\ \smallskip \\
22 October 2004
} 
\begin{document}
\maketitle%
\ \vspace{-5ex} \\
\begin{abstract}      
\noindent
The spectrum of the Laplace operator in a curved strip
of constant width built along an infinite plane curve,
subject to three different types of boundary conditions
(Dirichlet, Neumann and a combination of these ones, respectively),
is investigated.
We prove that the essential spectrum as a set is stable
under any curvature of the reference curve which vanishes at infinity
and find various sufficient conditions which guarantee
the existence of geometrically induced discrete spectrum.
Furthermore, we derive a lower bound to the distance
between the essential spectrum and the spectral threshold
for locally curved strips.
The paper is also intended as an overview of some new and old results
on spectral properties of curved quantum waveguides.
\end{abstract}

\newpage
{\small \tableofcontents}

\section{Introduction}
Let~$\Omega$ be a region (\ie\/ open connected set) in~$\Real^n$, $n \geq 1$,
with sufficiently regular boundary~$\partial\strip$,
and consider the corresponding Laplacian~$-\Delta$ on~$\sii(\Omega)$
with mixed Dirichlet-Neumann boundary conditions.
If~$\Omega$ is bounded, then it is well known that the spectrum
of the Laplacian is purely discrete, and properties of the eigenvalues
have been intensively studied.
On the other hand, it is easy to see that the spectrum is~$[0,\infty)$,
\ie~purely essential, if~$\Omega$ is unbounded and
sufficiently extended at infinity
(namely, it contains arbitrarily large balls).
Although it was shown already by F.~Rellich in 1948~\cite{Rellich}
that there exist unbounded regions whose spectrum contains
discrete eigenvalues (or it is even purely discrete!),
the spectral theory for the eigenvalues
has attracted much less attention than in the bounded case.

However, recent advent of mesoscopic physics has given
a fresh impetus to study the (discrete) spectrum of the Laplacian
in unbounded regions.
For, let us recall that the quantum Hamiltonian~$H$ of a free spin-less
particle of effective mass~$m^*$ constrained to a spatial region~$\Omega$,
\ie~$H=-\hbar^2/(2m^*)\Delta$ on~$\sii(\Omega)$,
represents a reasonable mathematical model for the dynamics
in various semiconductor structures devised and produced
in the laboratory nowadays.
Here it is mostly natural to consider the Dirichlet boundary conditions
on~$\partial\Omega$ corresponding to a large chemical potential barrier,
however, other situations modelling the impenetrable walls of~$\Omega$
(in the sense that there is no probability current through the boundary)
may be relevant as well (see \eg~\cite{KZ1,KZ2})
and can in principle model different types of inter\-phase in a solid.
We refer to~\cite{DE,LCM,Hurt}
for the physical background and references.
An important category of these systems is represented
by so-called \emph{quantum waveguides} which are modelled
by infinitely stretched tubular regions in~$\Real^n$ with~$n=2,3$.

The simplest situation occurs
if~$\Omega$ is an infinite plane strip,
\ie\/ a tubular neighbourhood of constant width along
an infinite curve in~$\Real^2$.
In~1989, P.~Exner and P.~\v{S}eba~\cite{ES} demonstrated the existence
of discrete spectrum for the Dirichlet Laplacian
in curved strips which were asymptotically straight and sufficiently thin.
Numerous subsequent studies
improved their result and generalized it to space tubes
\cite{GJ,RB,DE}.
For more information and other spectral and scattering
properties, see the review paper~\cite{DE} and references therein.
An important improvement was made by J.~Goldstone
and R.~L.~Jaffe in 1992~\cite{GJ}; the authors introduced
a variational argument which enables them to demonstrate
the existence of discrete eigenvalues without the restriction
on the width of the strip.
The paper~\cite{K1} deals with a more general situation
where the strip is not constructed in~$\Real^2$
but in a two-dimensional Riemannian manifold.
The evidently more complicated case of layers,
\ie~$\Omega$ is a tubular neighbourhood
about a complete non-compact surface in~$\Real^3$,
was investigated in~\cite{DEK1,DEK2,EK3,CEK,LL1}.

A common property of the Dirichlet systems cited above
is that a bending of a straight strip or layer generates
discrete eigenvalues below the essential spectrum,
\ie\/ geometrically induced quantum \emph{bound states},
which are known to disturb the particle transport.
The result is also interesting from the semiclassical point of view
because there are no classical closed trajectories in the tubes in question,
apart from a zero measure set of initial conditions in the phase space.
Hence, this is a pure quantum effect of geometrical origin.

On the mathematical side, the results are of interest
because the tubular neighbourhoods represent
a class of so-called \emph{quasi-cylindrical} regions,
for which the existence of discrete spectrum is a non-trivial property.
We refer to the books~\cite{Glazman} and~\cite{Edmunds-Evans}
for a classification of Euclidean regions
and basic properties of the spectrum of the Dirichlet Laplacian
as related to the form of an unbounded region.

The spectral results become richer if one considers
a combination of Dirichlet and Neumann boundary conditions~\cite{DKriz1,DKriz2}.
Here the problem is interesting even for straight strips
and much less studied in the literature.

Apart from the curved quantum waveguides, the discrete spectrum can
be also generated by a local deformation of the
boundary~$\partial\Omega$ of straight tubes and layers \cite{BGRS,BEGK,EV2},
via introducing an obstacle \cite{ELV,DP,APV}
or impurities modelled by a Dirac interaction \cite{EGST,EK1,EK2,EN},
coupling several waveguides by a window \cite{ESTV,EV,EV1,BE}, \etc.
The spectrum of periodically and randomly curved waveguides
was investigated in~\cite{Y,SS} and~\cite{KS}, respectively.
Finally, let us mention systems where~$\Omega=\Real^n$, $n=2,3$,
and the quantum waveguide is introduced by means of
a magnetic field \cite{EJKov1,EKov,EJKov2}
or a strong Dirac interaction supported by an infinite curve or surface 
\cite{EI,EY1,ESylwia1,ESylwia2,E4,ESylwia3,ESylwia4,ESylwia5,ESylwia6,EN1}.

The present paper is devoted to a study of the interplay
between the geometry, boundary conditions
(we consider uniform Dirichlet, Neumann or a combination of these ones)
and the spectral properties of the Laplacian in the infinite planar curved strips.
We referred above to the theory of quantum waveguides
as our main physical motivation.
Let us conclude this section by mentioning other fields in physics
where this might be a reasonable mathematical model.

Considerable current interest in designing integrated optoelectronic circuits
involves (classical) \emph{electromagnetic waveguides}
as a set of essential components.
In two-dimensional structures, planar symmetry implies that the
waveguide modes can have transversally electric,
respectively transversally magnetic
polarizations corresponding to Dirichlet,
respectively Neumann boundary conditions, \cf~\cite{MFJ}.
Maxwell's equations yield the similar spectral problem as above,
the difference is only in the physical meaning of the spectral parameter.

The eigenvalues of the Neumann Laplacian
may be also regarded as velocity potentials of an inviscid, irrotational
fluid or trapped vibrational modes of an~\emph{acoustic waveguide}.
We refer to~\cite{ELV,DP} for the considerable applied literature
on such problems.

Combinations of Dirichlet and Neumann boundary conditions appear
as a natural generalization of the uniform boundary conditions.
A physical system satisfying such a combination is the
\emph{Earth-ionosphere waveguide}: it is known that for very low frequencies
the electromagnetic wave dynamics between the Earth and the ionosphere
can be approximated as a propagation between the plates
with the perfect electric (the Earth) and perfect magnetic (the ionosphere) conductors,
see~\cite{OM} and references therein.
Problems of this type arises naturally in many other areas of physics,
most notably in theoretical studies of superconductivity, photonic
crystals, \etc.

\section{Scope of the Paper}\label{Sec.Scope}
\begin{figure}[ht]
\begin{center}
\epsfig{file=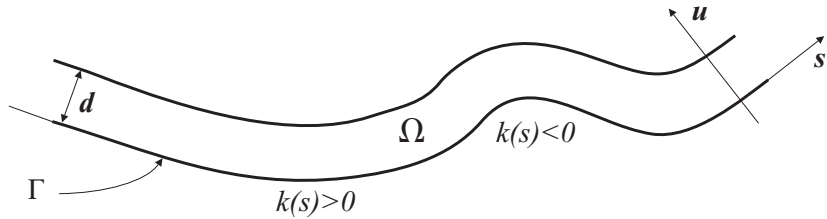,width=0.9\textwidth}
\end{center}
\caption{
Configuration space~$\strip$ defined as a strip
over an infinite curve~$\curve$ in~$\Real^2$.
}\label{Figure1}
\end{figure}

The main aim of the present paper is to study the geometrically
induced (discrete) spectrum of the operator~$H^\both$
defined as the Laplacian $-\Delta$ on $\sii(\strip)$,
where~$\strip$ is a (one-sided) tubular neighbourhood of a fixed width $d>0$
along an infinite plane curve~$\curve$ of curvature~$k$,
see~Figure~\ref{Figure1}.
We adopt suitable hypotheses (\cf~$\Hi$ below) in order to ensure that
the boundary $\partial\strip$ consists of two parallel connected embedded curves
of class~$\Smooth^2$.
The index~$\both$ will distinguish three
different types of boundary conditions considered here.
Namely, we consider the recently widely investigated
\emph{Dirichlet} boundary condition ($\both:=D$),
the \emph{Neumann} boundary condition ($\both:=N$)
and the simplest \emph{combination} of the both just mentioned ($\both:=DN$):
the Dirichlet boundary condition imposed
on one connected component of~$\partial\strip$
and the Neumann condition on the other one.%

If the reference curve~$\curve$ is a \emph{straight} line,
then it is rather a textbook exercise
to analyse the operator~$H^\both$ by means of a separation of variables
and conclude that its spectrum is purely absolutely continuous
and equals the interval $[E_1^\both,\infty)$,
where the non-negative value~$E_1^\both$
is determined by the respective boundary conditions,
\cf~(\ref{TransEV}).
However, the spectral problem for~$H^\both$
becomes always difficult whenever~$\curve$ is \emph{curved},
and two basic questions arise in this context:
\begin{enumerate}
\item
Which geometry preserves the essential spectrum $[E_1^\both,\infty)$?
\item
Which geometry produces a spectrum below $E_1^\both$?
\end{enumerate}
These questions represent ultimate concern of this paper.
We try to make a survey of known answers and
contribute to the problem by our own results.
Furthermore, if the spectrum below~$E_1^\both$ exists,
we establish various estimates of the spectral
threshold $\inf\sigma(H^\both)$.
It should be stressed here that the existence of \emph{discrete} spectrum,
\ie~the issue mentioned in Introduction,
is proved whenever the considered geometry
is in accordance with \emph{both} the above questions
(because then the spectrum below~$E_1^\both$
consists of isolated eigenvalues of finite multiplicity only).

Concerning the first question,
we show that the essential spectrum of a curved strip
coincides with the spectrum of the straight one
provided the reference curve~$\curve$ is straight asymptotically
in the sense that its curvature vanishes at infinity,
\cf~Theorem~\ref{Thm.Ess}.
Although this sufficient condition is very natural
and in perfect accordance with the intuition,
it is for the first time in this paper when
the essential spectrum is localized
without imposing any additional conditions
(\eg, about the decay of the derivatives of curvature
at infinity, \cf~\cite{DE,RB,DKriz2}).
The progress has become possible due to
a general characterization of essential spectrum adopted
from a paper by Y.~Dermenjian \etal~\cite{DDI} (\cf~our Lemma~\ref{Iftimie}),
which is for our purposes more suitable
than the classical Weyl criterion.
On the other hand, periodic strips are discussed
as an illustration of asymptotically non-straight geometry
which does change the essential spectrum.

The answer to the second question depends substantially
on the choice of boundary conditions.
First of all, notice that the question does not make sense
for the Neumann strips because $E_1^N=0$,
\cf~Theorem~\ref{thm.N}.
A characteristic property of the Dirichlet strips
is that any bending of the reference curve~$\curve$
pushes the infimum of the spectrum
below the spectral threshold~$E_1^D>0$
of the corresponding straight strip,
\cf~Theorem~\ref{Thm.Disc.Dirichlet}.
This property was shown first in~\cite{ES} for sufficiently thin strips
and the proof for more general cases was introduced in~\cite{GJ}.
On the other hand, the case of combined Dirichlet-Neumann boundary
conditions was introduced quite recently in~\cite{DKriz2}.
The authors established the existence of spectrum below $E_1^{DN}>0$
provided the total bending angle of~$\curve$
(\ie\ the integral of curvature, \cf~(\ref{bending angle}))
has a suitable sign.
In this paper, we generalize this result
and add two new sufficient conditions, \cf~Theorem~\ref{thm.DN}.
We also derive an interesting result on the number
of discrete eigenvalues of~$H^{DN}$, \cf~Proposition~\ref{Prop.Number}

Finally, when~$\curve$ is curved only locally,
we derive an upper bound to the spectral threshold,
\ie~$\inf\sigma(H^\both)$,
for the Dirichlet strip
and the one with combined Dirichlet-Neumann boundary condition,
\cf~Theorem~\ref{thm.estimate}.
In particular, we find important qualitative differences
between these two respective results.
Making the curvature small,
the leading term in the estimate of the difference
$\inf\sigma(H^D)-E_1^D$ is proportional to the fourth power
of the total bending angle,
while it is the second power what one obtains
for the Dirichlet-Neumann case,
\cf~Remark~\ref{Rem.mildly}.
Another interesting difference appears when we are shrinking the
width of the strip to zero, \cf~Remark~\ref{Rem.Thin}.
These estimates are new in the theory of curved quantum waveguides.
We can only compare them with the eigenvalue asymptotics
for mildly curved, respectively thin, Dirichlet strips
established in~\cite{DE}.
Let us note that a similar estimate for straight, window-coupled waveguides
was given in~\cite{EV,EV1}, see also~\cite{BEG}.

All our proofs of the statements concerning the existence
and properties of the spectrum below~$E_1^\both$
are based on a variational strategy.
The corner stone of them, \ie~the construction of a suitable
trial function, follows the idea of~\cite{GJ}, see also~\cite{DE,RB}.

The paper is organized as follows.
Section~\ref{Sec.Prel} is devoted to some preliminary material
in order to be able to state precisely the main results
of the paper, \ie~Theorems~\ref{Thm.Ess}--\ref{thm.estimate},
in the subsequent Section~\ref{Sec.Results}.
The proofs and discussions of the Theorems are presented
in Sections~\ref{Sec.Ess} and~\ref{Sec.Dis}.
We conclude the paper by Section~\ref{Sec.Conclusions}, where some
open problems and directions of a future research are mentioned.

\section{Preliminaries}\label{Sec.Prel}
\subsection{Configuration space}
Let~$\curve$ be a unit-speed infinite plane curve,
\ie~the (image of the) $\Smooth^2$-smooth embedding 
$
  \curve: \Real\to\Real^2 :
  \big\{s\mapsto\big(\Gamma^1(s),\Gamma^2(s)\big)\big\}
$
satisfying $|\dot{\curve}(s)|=1$ for all $s\in\Real$
(the arc-length parameter of the curve).
The function $N:=(-\dot{\curve}^2,\dot{\curve}^1)$ defines
a unit normal vector field and the couple $(\dot{\curve},N)$
gives a distinguished Frenet frame, \cf~\cite[Chap.~1]{Kli}.
The curvature is defined through the Frenet-Serret formulae by
$k:=\det(\dot{\curve},\ddot{\curve})$.
We note that~$k$ is a continuous function
of the arc-length parameter
and the sign of~$k(s)$ is defined uniquely
up to the re-parameterization $s \mapsto -s$.
It is also worth to notice that the curve~$\curve$ is fully determined
(except for its position and orientation in the plane)
by the curvature function~$k$ only, \cf~\cite[Sec.~II.~20]{Kreyszig}.

Let~$d>0$, $\trans:=(0,d)$ and
$\strip_0:=\Real\times\trans$ be a straight strip of width~$d$.
We define a curved strip of the same width based on~$\curve$
via~$\strip:=\stripmap(\strip_0)$, where
\begin{equation}\label{StripMap}
  \stripmap: \Real^2\to\Real^2:
  \left\{ (s,u) \mapsto \curve(s)+u\,N(s) \right\} .
\end{equation}
Through all the paper, we always assume that
\begin{Assumption}{\Hi}
  $\strip$ is not self-intersecting
  \quad and \quad
  $k\in\sinf(\Real)$
  \ with \
  $d\,\|k_+\|_\infty<1$,
\end{Assumption}
where~$k_\pm:=\max\{0,\pm k\}$.
Then $s\mapsto\stripmap(s,u)$ for~$u\in\overline{I}$ fixed
traces out a parallel curve at a distance~$|u|$ from~$\curve$
and $u\mapsto\stripmap(s,u)$ for~$s\in\Real$ fixed
is a straight line orthogonal to~$\curve$ at~$s$.
Furthermore, the mapping $\stripmap: \strip_0 \to \strip$
is a $\Smooth^1$-diffeo\-morphism and its inverse determines
a system of natural ``coordinates''~$(s,u)$ in a neighbourhood of~$\curve$.
We remark that under our assumption~$\Hi$
the curve $\stripmap(\Real \times \{u\})$ is of class~$\Smooth^2$
for any fixed $u \in \overline I$,
in particular, this claim holds true for both the boundary curves.
\begin{rem}
In this paper, we adopt the standard component notation of tensor analysis
together with the repeated indices convention.
The range of indices is~$1,2$
and they are associated with the above mentioned coordinates
via~$(1,2)\leftrightarrow(s,u)$.
The partial derivatives are marked by a comma with the index.
\end{rem}

By virtue of the Frenet-Serret formulae,
the metric tensor of~$\strip$ in these coordinates,
\ie~\mbox{$G_{ij}:=\stripmap_{,i}\cdot\stripmap_{,j}$}
where~``$\cdot$'' denotes the scalar product in~$\Real^2$,
has the following diagonal form
\begin{equation}\label{metric}
  \big(G_{ij}(s,u)\big)=
  \begin{pmatrix}
    \left(1-u k(s)\right)^2 & 0 \\
    0        & 1
  \end{pmatrix}.
\end{equation}
Its determinant, $G:=\det(G_{ij})$, defines through
$d\strip:=G(s,u)^\demi ds du$ the area element of the strip.
By virtue of the second part of the assumption~$\Hi$,
it is clear that the metric~(\ref{metric}) is uniformly
elliptic. In particular, we have the following useful estimates:
\begin{equation}\label{1<G<1}
  \forall (s,u)\in\strip_0 \:: \quad
  C_-
  \leq 1-uk(s) \leq
  C_+
  \qquad\textrm{with}\qquad
  C_\pm:=1\pm d\,\|k_\mp\|_\infty .
\end{equation}
%

\subsection{The Laplacian}
Our object of interest is the Laplacian $-\Delta$ on
$\sii(\Omega)$, subject to various boundary conditions imposed
on~$\partial\strip$. Our basic strategy is to use the
diffeomorphism $\stripmap: \strip_0 \to \strip$ in order to
replace the simple operator~$-\Delta$ on the complicated Hilbert
space~$\sii(\Omega)$ by a more complicated operator~$H^\both$ on
the simpler Hilbert space~$\Hilbert := \sii(\Omega_0, d\strip)$.
In particular, $H^D$ is the operator replacing the Laplacian with
Dirichlet boundary condition, $H^N$ corresponds to the Neumann
boundary condition and $H^{DN}$ has the Dirichlet boundary
condition imposed on the reference
curve~$\curve\equiv\stripmap(\Real\times\{0\})$ and the Neumann one imposed on
the opposite boundary $\stripmap(\Real\times\{d\})$. Sometimes, we
shall use the common superscript~$\both\in\{D,N,DN\}$ to consider
two or all of the three different situations simultaneously.

More precisely, the operators~$H^\both$
are introduced as the unique self-adjoint operators
associated on~$\Hilbert$ with the quadratic forms~$Q^\both$ defined by
\begin{align}
  Q^\both[\psi] &:=
  \left( \psi_{,i},G^{ij}\psi_{,j} \right) , \label{Dform}
  \\
  \Dom Q^D &:=\sobi(\strip_0,d\strip) , \label{Dom.Dform}
  \\
  \Dom Q^N &:=\Sobi(\strip_0,d\strip) ,  \label{Nform}
  \\
  \Dom Q^{DN} &:=\left\{\psi\in\Sobi(\strip_0,d\strip) \:| \
  \psi(s,0)=0 \quad \textrm{for a.e.} \ s \in\Real \right\} .\label{DNform}
\end{align}
Here and in what follows,
$(G^{ij})$ stands for the inverse of~$(G_{ij})$,
$(\cdot,\cdot)$ and~$\|\cdot\|$  denotes the scalar product
and the norm in~$\Hilbert$, respectively,
and $\psi(\cdot,0)$ means the trace of the function~$\psi$
on the boundary part~$\stripmap(\Real\times\{0\})$.
\begin{rem}\label{RemNormEquiv}
Since the metric $(G_{ij})$ is uniformly elliptic due
to~(\ref{1<G<1}), it is not necessary to take into account the
measure~$d\Omega$ in~(\ref{Dom.Dform}), (\ref{Nform})
and~(\ref{DNform}).
\end{rem}
\begin{rem}[Operators associated with~$Q^\both$]
We have
\begin{equation}\label{Hamiltonian}
  H^\both
  =-G^{-\demi}\partial_i G^\demi G^{ij} \partial_j
  \,,
\end{equation}
which is a general expression for the Laplacian
in a manifold equipped with a metric~$(G_{ij})$.
The equality in~(\ref{Hamiltonian}) must be understood
in the form sense if the curvature~$k$ is not differentiable
(which is the case we are particularly concerned to deal with
in this paper).
Nevertheless, assuming that the reference curve~$\curve$
is, say, $\Smooth^3$-smooth, then the metric is differentiable
and, putting~(\ref{metric}) into~(\ref{Hamiltonian}),
we can write
\begin{equation*}
  H^\both
  =-\frac{1}{(1-uk(s))^2}\,\partial_s^2
  -\frac{u\dot{k}(s)}{(1-u k(s))^3}\,\partial_s
  -\partial_u^2
  +\frac{k(s)}{1-u k(s)}\,\partial_u
\end{equation*}
as an operator identity on the functions from
$
  \Dom H^\both
$.
Moreover, the operator domain~$\Dom H^\both$ is exactly that
subset of the space~$\Sobii(\strip_0)$ whose elements satisfy the
corresponding boundary conditions on~$\partial \strip_0$ in the
classical sense, \cf~\cite{Kriz}.
\end{rem}
%

\subsection{Straight strips}\label{Sec.Straight}
If the strip is straight in the sense that $k \equiv 0$,
\ie~$k$ is equal to zero everywhere on~$\Real$,
then the Laplacian coincides with the decoupled operator
\begin{equation}\label{Hamiltonian.straight}
  H^\both_0 :=
  \overline{-\Delta^\Real \otimes \Id + \Id \otimes (-\Delta_\both^\trans)}
  \qquad\textrm{on}\quad\
  \sii(\Real)\otimes\sii(\trans),
\end{equation}
where~$\Id$ denotes the identity operator on appropriate spaces.
The operators on the transverse section, $-\Delta_\both^\trans$,
are the usual Laplacians on~$\sii(\trans)$ with the Dirichlet
boundary conditions if~$\both=D$,
the Neumann conditions if~$\both=N$,
or the Dirichlet condition at~$0$
and the Neumann one at~$d$ if~$\both=DN$.
The eigenvalues of~$-\Delta_\both^\trans$ are given by
\begin{equation}\label{TransEV}
  E_n^D:=(\pi/d)^2 n^2,
  \quad
  E_n^{N}:=(\pi/d)^2 (n-1)^2,
  \quad
  E_n^{DN}:=(\pi/d)^2 (n-\mbox{$\demi$})^2,
\end{equation}
where $n\in\Nat\!\setminus\!\{0\}$.
The corresponding family of normalized
eigenfunctions $\{\chi_n^\both\}_{n=1}^\infty$
can be chosen in the following way:
\begin{align}
  \chi_n^\both(u) &:=
  \sqrt{\mbox{$\frac{2}{d}$}}\,\sin\sqrt{E_n^\both} \, u
  \quad\ \textrm{for}\quad \both\in\{D,DN\} \,;
  \label{TransEF} \\
  \chi_n^N(u) &:=
  \begin{cases}
    \sqrt{\mbox{$\frac{1}{d}$}}
    & \ \textrm{if} \quad n=1 ,
    \label{TransEF.N} \\
    \sqrt{\mbox{$\frac{2}{d}$}}\,\cos\sqrt{E_n^N} \, u
    & \ \textrm{if} \quad n\geq 2 .
  \end{cases}
\end{align}
In view of~(\ref{Hamiltonian.straight}) and~\cite[Thm.~VIII.33]{RS1},
the straight strip has an absolutely continuous spectrum
starting from the first eigenvalue of the transverse Laplacian, \ie,
\begin{equation}\label{StraightSpectrum}
  \sigma(H_0^\both)=\sigma_\mathrm{ess}(H_0^\both)
  =[E_1^\both,\infty) .
\end{equation}
%

\section{Main results}\label{Sec.Results}
As we have seen, the essential spectrum of a straight strip,
\ie~$k \equiv 0$, is the interval $[E_1^\both,\infty)$.
In Section~\ref{Sec.Ess}, we prove that the same spectral result
holds for any curved strip which is straight
\emph{asymptotically} in the sense that the curvature~$k$ vanishes
at infinity, \ie,
\begin{Assumption}{\di}
  $k(s) \xrightarrow[|s|\to\infty]{} 0$ .
\end{Assumption}
\begin{thm}[Essential spectrum]\label{Thm.Ess}
Suppose~$\Hi$. If the strip satisfies~$\di$, then
$$
  \sigma_\mathrm{ess}(H^\both) = [E_1^\both,\infty)
  \qquad\textrm{for}\quad\both\in\{D,N,DN\}.
$$
\end{thm}

To the best of our knowledge, the spectrum of the Neumann Laplacian~$H^N$,
has been previously investigated just for strips which were straight
and contained an obstacle, \cite{ELV,DP}.
Hence, our Theorem~\ref{Thm.Ess} represents
a quite new result concerning the spectral theory of curved Neumann strips.

The Dirichlet-Neumann case, \ie~$\both=DN$, was previously considered
just in the recent letter~\cite{DKriz2}. It is mentioned there
that $\inf\sigma_\mathrm{ess}(H^{DN})=E_1^{DN}$ provided~$k$
has a compact support. Here we have proved that
the whole interval $[E_1^{DN},\infty)$ is in the essential spectrum
under much weaker condition~$\di$.

Although the case of Dirichlet strips, \ie~$\both=D$, has already
been considered in many works, our Theorem~\ref{Thm.Ess}
represents a new result in this situation as well, since
it is for the first time when the whole essential spectrum
has been localized under a condition which does not contain
derivatives of~$k$.
Some decay assumptions about the derivatives
of the curvature were even required in order to localize
the threshold $\inf\sigma_\mathrm{ess}(H^D)$ itself
in the previous works, \cf~\cite{DE,RB}.
(An exception is the paper~\cite{K1}
where, however, only a lower bound on the threshold is given.)
Let us mention that the result of Theorem~\ref{Thm.Ess}
was achieved in the thesis~\cite{these} under an additional condition
about vanishing of the first derivative of~$k$.

Since~$H^N$ is non-negative,
it follows immediately from Theorem~\ref{Thm.Ess}
that there is no discrete spectrum
in asymptotically straight Neumann strips.
\begin{thm}[Neumann case]\label{thm.N}
Suppose~$\Hi$. Then
\begin{itemize}
\item[]
$
  \inf\sigma(H^N) = E_1^N \equiv 0 .
$
\end{itemize}
Consequently,
if the strip is asymptotically straight, \ie~$\di$,
then
$$
  \sigma(H^N) = \sigma_\mathrm{ess}(H^N) = [0,\infty),
$$
\ie, $\sigma_\mathrm{disc}(H^N) =\emptyset$.
\end{thm}
\noindent
Here the fact that the spectral threshold of~$H^N$
starts exactly at~$0$ for any strip
can be easily proved by means of a suitable trial function
(\cf~Proposition~\ref{Prop.upperbound}).

An interesting result in the theory of quantum waveguides
is that the curved geometry may produce a non-trivial spectrum
below the energy~$E_1^\both$ for $\both\in\{D,DN\}$.
The phenomenon is examined in this paper.
Notice that any result of the type $\inf\sigma(H^\both)< E_1^\both$
together with the decay condition~$\di$ yield
that the spectrum below~$E_1^\both$ consists of isolated
eigenvalues of finite multiplicity only,
\ie~$\sigma_\mathrm{disc}(H^\both)\not=\emptyset$.
However, we do not restrict ourselves to the particular
case of asymptotically straight strips, \ie,
the geometrically induced spectrum below~$E_1^\both$
may have a non-zero Lebesgue measure, too.

Sufficient conditions for the Laplacians $H^\both$ with~$\both\in\{D,DN\}$
to have a non-empty spectrum below $E_1^\both$ are known.
In particular, any non-trivial curvature of the reference
curve pushes the spectrum of~$H^D$ down the corresponding
spectral threshold of the straight strip.
\begin{thm}[Dirichlet case] \label{Thm.Disc.Dirichlet}
Suppose~$\Hi$.
\begin{itemize}
\item[]
If \ $k \not\equiv 0$, then \
$
  \inf\sigma(H^D) <  E_1^D .
$
\end{itemize}
Consequently, if the strip is not straight
but it is straight asymptotically, \ie~$\di$,
then~$H^D$ has at least one eigenvalue of finite multiplicity
below its essential spectrum $[E_1^D,\infty)$,
\ie, $\sigma_\mathrm{disc}(H^D)\not=\emptyset$.
\end{thm}
\noindent
This property was shown first in~\cite{ES}
for sufficiently thin strips with a rapidly decaying curvature
and since various improvements have been achieved
(see the references mentioned in Introduction, mainly~\cite{GJ}).
We find useful to make a proof of Theorem~\ref{Thm.Disc.Dirichlet}
in Section~\ref{Sec.Existence} since it can be made
simultaneously with the proof of the new result contained
in condition~(a) of Theorem~\ref{thm.DN} below.

As for the operator~$H^{DN}$, its spectrum was studied
for the first time in the recent letter~\cite{DKriz2}.
It shows that the position of the infimum of spectrum essentially
depends on the sign of the total bending angle
\begin{equation}\label{bending angle}
  \alpha\,:=\,\int_\Real k(s)\,ds ,
\end{equation}
which is well defined if we assume that the curvature is integrable.
In detail, the authors of~\cite{DKriz2}
proved that: i)~the spectrum of~$H^{DN}$ in
a non-trivially curved strip starts below~$E_1^{DN}$
provided~$\alpha \leq 0$ and the curvature~$k$
is non-positive out of some bounded interval.
On the other hand, ii)~if~$k(s) \geq 0$ for all $s\in\Real$,
then the spectrum below the energy~$E_1^{DN}$ is empty.
Our improvement is two-fold.
Firstly, we generalize the first claim in the sense
that we skip the condition on~$k$.
Secondly, we find a sufficient condition which guarantees
the existence of spectrum below~$E_1^{DN}$
even for some strips with~$\alpha>0$.
In addition to these substantial generalizations,
we will derive the same result also for periodic waveguides.
Let us summarize the spectral properties of~$H^{DN}$
into the following theorem.

\begin{figure}[h]
\begin{center}
\epsfig{file=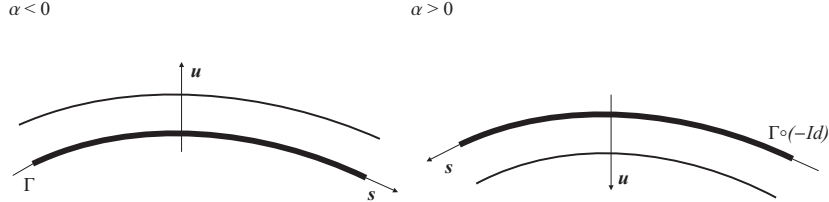,width=0.9\textwidth}
\end{center}
\caption{Inversion of orientation of the reference curve
(given by the re-parameterization~$s\mapsto -s$).
Thick lines denote the Dirichlet boundary condition, thin lines
the Neumann one.
}
\label{Figure2}
\end{figure}

\pagebreak[4]%
\begin{thm}[Dirichlet-Neumann case]\label{thm.DN}
Suppose~$\Hi$.%
\begin{itemize}%
\item[\emph{(i)}]
If \ $k \not\equiv 0$, then any of the three conditions
\begin{itemize}
\item[\emph{(a)}]\
$k\in\si(\Real)$ \ and \ $\alpha \equiv \int_\Real k(s) \, d s\leq 0$
\item[\emph{(b)}]\
$k$ is periodic
\item[\emph{(c)}]\
$k_- \not\equiv 0$ \ and \ $d$ is small enough
\end{itemize}
is sufficient to guarantee that \
$
  \inf\sigma(H^{DN}) <  E_1^{DN} .
$
\item[\emph{(ii)}]
If \ $k_- \equiv 0$, then \
$
  \inf\sigma(H^{DN}) \geq  E_1^{DN} .
$
\end{itemize}
Consequently, if the strip is not straight
but it is straight asymptotically, \ie~$\di$,
then any of the conditions~\emph{(a)} or~\emph{(c)} is sufficient to guarantee
that~$H^{DN}$ has at least one eigenvalue of finite multiplicity
below its essential spectrum $[E_1^{DN},\infty)$,
\ie, $\sigma_\mathrm{disc}(H^{DN})\not=\emptyset$.
On the other hand, if the strip is asymptotically straight
and~$k_- \equiv 0$, then
$\sigma(H^{DN})=\sigma_\mathrm{ess}(H^{DN})= [E_1^{DN},\infty)$,
\ie, $\sigma_\mathrm{disc}(H^{DN})=\emptyset$.
\end{thm}
\begin{rem}
The signs of~$k(s)$ and the corresponding total bending angle~$\alpha$
change after the change of arc-length parameter given by~$s \mapsto -s$.
It has to be stressed here that such a re-parameterization
of the reference curve~$\curve$
leads to another strip due to~(\ref{StripMap}) and,
consequently, there is no ambiguity in stating
the spectral results on~$H^{DN}$ in terms of the sign
of~$\alpha$ and~$k$, see Figure~\ref{Figure2}.
\end{rem}
\noindent
The sufficient conditions (a)--(c) of the first part of Theorem~\ref{thm.DN}
are proved in Section~\ref{Sec.Existence}. 
We refer to~\cite{DKriz2} for the original proof of the part~(ii)
(this proof is in fact very technical,
based on a decomposition of~$H^{DN}$ to the transverse basis~(\ref{TransEF})
and the spectral analysis of an associated ordinary differential operator)
and to~\cite{FK} for a recent, simplified proof.
A comparison of the condition~(a) with the assumptions given in~\cite{DKriz2}
is done in Remark~\ref{Rem.comparison}.

\medskip
Consider now a situation when the discrete
spectrum of~$H^\both$, $\both\in\{D,DN\}$,
below the energy~$E_1^\both$ is not empty.

Although this paper in not intended to investigate
the \emph{number} of eigenvalues of~$H^\both$,
let us point out the following remarkable property of~$H^{DN}$
which we establish at the end of Section~\ref{Sec.Existence}.
\begin{prp}[Number of bound states in the DN case]\label{Prop.Number}
Suppose~$\Hi$ and~$\di$. If $k_- \not \equiv 0$ then
$$
  \forall n \in \Nat \quad \exists d_n > 0\ : \quad
  d < d_n \ \Longrightarrow \ N(H^{DN})
\geq n\,,
$$
where~$N(H^{DN})$ denotes the number of discrete eigenvalues
of~$H^{DN}$, counting multiplicity.
\end{prp}
\noindent
The number of bound states in thin strips is another property,
which demonstrates a significant influence of the choice of boundary conditions
on the spectrum.
To see it, we recall that an upper SKN-type (\cf~\cite{Seto,Kl,Newton}) bound
on the number of bound states in thin Dirichlet strips
was derived in~\cite[Sec.~2.3]{DE}
and it showed that~$N(H^D)$ is bounded from above
by a finite constant which \emph{does not} depend on the strip width~$d$.
On the other hand, Proposition~\ref{Prop.Number} shows that~$N(H^{DN})$
can reach arbitrarily large value by shrinking the strip width to zero.

The last objective of this paper is to estimate the distance
between the bottom of the essential spectrum~$E_1^\both$ and
the spectral threshold~$\inf\sigma(H^\both)$
(which will represent the lowest eigenvalue since,
for this problem, we restrict ourselves
to the strips with curvature having compact support).
We derive the following upper bounds (to the lowest eigenvalue),
which are again qualitatively different for the Dirichlet
and mixed Dirichlet-Neumann situations, respectively.

\begin{thm}[Estimates of the spectral threshold]\label{thm.estimate}
Suppose $\Hi$ and assume that~$k$ has a compact support
in an interval of width $2s_0$.
\begin{itemize}
\item[\emph{(i)}]
If \ $\alpha \leq 0$, then \
$
  \inf \sigma(H^{DN})\,\leq\,E_1^{DN}
   - {C^{DN}(s_0,d,\alpha)}^2\,\alpha^2\,,
$
where
\begin{equation*}
  C^{DN}(s_0,d,\alpha)
:= \sqrt{E_1^{DN}}\,
   \frac{\sqrt{3}/\pi}{1+\sqrt{1-\frac{3}{2}\frac{\alpha s_0}{d}
   + \frac{3}{4}\alpha^2 \left(\frac{1}{2} +
   \frac{2}{\pi^2}\right)}} \,.
\end{equation*}
\item[\emph{(ii)}]
$
  \inf \sigma(H^{D})\,\leq\,E_1^{D}
   -{C^D(s_0,d,\alpha)}^2\,\alpha^4
  \,,
$
where
\begin{align*}
   C^D(s_0,d,\alpha)
:= \frac{2^4}{3^3}\,
\frac{\sqrt{3}/\pi^2}{d\left( \frac{s_0}{d} - \frac{\alpha}{4} +
\frac{2}{3\,\pi} \right)}\,
\frac{1}{1 + \sqrt{1 + \left(\frac{4\,\alpha}{3\,\pi}\right)^2
\frac{4 s_0 - \alpha\,d}
{4 s_0 - \alpha\,d +\frac{8 d}{3\,\pi}}}
}\,.
%
\end{align*}
\end{itemize}
\end{thm}
\noindent
These estimates are new in the theory of quantum waveguides and we
derive them in Section~\ref{Sec.Dis}.
One can immediately see that for small total bending angles, the
leading term in the estimate~(i) is proportional to the second power
of~$\alpha$, while it is the fourth power of~$\alpha$ in the
estimate~(ii).
Another essential difference in our estimates appears in the limit
case of thin strips.
We discuss these interesting disparities in Remarks~\ref{Rem.mildly}
and~\ref{Rem.Thin}.
We also compare there the result~(ii) with the exact eigenvalue
asymptotics obtained in~\cite{DE} by perturbation methods
applied to mildly curved or thin strips, respectively.

\section{Essential spectrum}\label{Sec.Ess}
This section is devoted to the proof of Theorem~\ref{Thm.Ess}.
It is achieved in two steps.
Firstly, in Lemma~\ref{Lemma.Ess1},
we employ a Neumann bracketing argument in order
to show that the threshold of the essential spectrum
does not descend below the energy~$E_1^\both$.
Secondly, in Lemma~\ref{Lemma.Ess2},
we prove that all energies above~$E_1^\both$
belong to the spectrum by means of the following general
characterization of essential spectrum which we have adopted
from~\cite{DDI}.
\begin{lmm}\label{Iftimie}
Let~$H$ be a non-negative self-adjoint operator
in a complex Hilbert space~$\Hilbert$ and $Q$ be the
associated quadratic form.
Then  $\eta\in\sigma_\mathrm{ess}(H)$ if and only if
$$
  \exists \{\psi_n\}_{n=1}^\infty\subset\Dom Q:\
  \left\{
  \begin{aligned}
    \emph{(i)} \quad
    & \forall n\in\Nat\setminus\{0\}:\ \|\psi_n\|=1, \\
    \emph{(ii)} \quad
    & \psi_n \xrightarrow[n\to\infty]{w} 0
    \quad\textrm{in}\ \Hilbert, \\
    \emph{(iii)} \quad
    & (H-\eta)\psi_n \xrightarrow[n\to\infty]{ } 0
    \quad\textrm{in}\ \left(\Dom Q\right)^*.
  \end{aligned}
  \right.
$$
\end{lmm}
\noindent
Here $\left(\Dom Q\right)^*$ denotes the dual of the space $\Dom Q$.
We note that
$
  H+1:\Dom Q\to\left(\Dom Q\right)^*
$
is an isomorphism and
\begin{equation}\label{-1norm}
  \|\psi\|_{-1} :=
  \|\psi\|_{\left(\Dom Q\right)^*} =
  \sup_{\phi\in\Dom Q\setminus\{0\}}
  \frac{|(\phi,\psi)|}{\|\phi\|_1} \,
\end{equation}
with
\begin{equation*}
\|\phi\|_1 := \sqrt{Q[\phi] + \|\phi\|^2}\,.
\end{equation*}

Lemma~\ref{Iftimie} is proved in a quite similar fashion
as the Weyl criterion, \cite[Thm.~7.24]{Weidmann}.
The advantage of the present characterization is that it requires
to find a sequence from the form domain of~$H$ only,
and not from~$\Dom H$ as it is required by the Weyl criterion.
Moreover, in order to check the limit from~(iii),
it is still sufficient to consider the operator~$H$ in the form sense,
\ie\ we will not need to assume that~$(G_{ij})$ is differentiable
in our case.

We start by an estimate on the threshold of the essential spectrum.
\begin{lmm}\label{Lemma.Ess1}
If~$\di$ holds true, then
$
  \inf\sigma_\mathrm{ess}(H^\both) \geq E_1^\both .
$
\end{lmm}
\begin{proof}
Since the curvature vanishes at infinity,
for any fixed~$\epsilon>0$, there exists~$s_\epsilon$ such that
\begin{equation}\label{estimates}
  \forall (s,u)\in\strip_\mathrm{ext} \,: \
  (1-d\epsilon)
  \leq
  1-u\,k(s)
  \leq
  (1+d\epsilon),
\end{equation}
where $\strip_\mathrm{ext}:=\strip_0\setminus\overline{\strip}_\mathrm{int}$
with $\strip_\mathrm{int}:=(-s_\epsilon,s_\epsilon)\times \trans$.
Denote by~$H_N^\both$ the operator~$H^\both$
with a supplementary Neumann boundary condition
imposed on the two segments~$\{\pm s_\epsilon\}\times\trans$, that is,
the operator associated with the form
$Q_N^\both:=Q_N^{\both,\mathrm{int}} \oplus Q_N^{\both,\mathrm{ext}}$,
where
\begin{align*}
  Q_N^{\both,\omega}[\psi]
 &:=  \left(\psi_{,i},
  G^{ij}\psi_{,j}\right)_{\sii(\strip_\omega,d\strip)},
  \\
  \Dom Q_N^{D,\omega}
 &:=  \left\{
  \psi\in\Sobi(\strip_\omega,d\strip)\,|\,
  \psi(s,0)=\psi(s,d)=0\ \mathrm{for\ a.e.}\
  s\in\Real \cap \overline \Omega_\omega
  \right\}\!, \\
  \Dom Q_N^{N,\omega}
 &:= \, \Sobi(\strip_\omega,d\strip)\, \\
  \Dom Q_N^{DN,\omega}
 &:=  \left\{
  \psi\in\Sobi(\strip_\omega,d\strip)\,|\,
  \psi(s,0)=0\ \mathrm{for\ a.e.}\
  s\in\Real \cap \overline \Omega_\omega
  \right\}
\end{align*}
for $\omega\in\{\mathrm{int},\mathrm{ext}\}$.
Since $H^\both \geq H_N^\both$
and the spectrum of the operator associated with~$Q_N^{\both,\mathrm{int}}$
is purely discrete, \cf~\cite[Chap.~7]{Davies},
the minimax principle gives the estimate
$$
  \inf\sigma_\mathrm{ess}(H^\both)
  \geq
  \inf\sigma_\mathrm{ess}(H_N^{\both,\mathrm{ext}})
  \geq
  \inf\sigma(H_N^{\both,\mathrm{ext}}),
$$
where~$H_N^{\both,\mathrm{ext}}$ denotes the operator
associated with~$Q_N^{\both,\mathrm{ext}}$.
Neglecting the non-negative ``longitudinal''
part of the Laplacian in~(\ref{Hamiltonian})
(\ie\ the term where one sums over~$i=j=1$)
and using the estimates~(\ref{estimates}),
we arrive easily at the following lower bound
$$
  H_N^{\both,\mathrm{ext}}
  \geq
  \frac{1-d\epsilon}{1+d\epsilon} \ E_1^\both
  \qquad\textrm{in}\quad
  \sii(\Omega_\mathrm{ext},d\strip),
$$
which holds in the form sense
(see also proof of Theorem~4.1 in~\cite{DEK2}).
The claim then follows by the fact that~$\epsilon$
can be chosen arbitrarily small.
\end{proof}
\begin{rem}[Neumann case]
Since $E_1^N=0$ and~$H^N$ is a non-negative operator,
the statement of Lemma~\ref{Lemma.Ess1} holds trivially true
for the Neumann boundary conditions, \ie, $\both=N$,
even without the assumption~$\di$.
\end{rem}
\begin{exa}[Periodic waveguides]\label{Example.Ess.D}
The periodic strip (\ie\/ as\-sump\-tion~$\di$ is not obeyed)
is the simplest example for which
$$
  \inf \sigma_\mathrm{ess}(H^\both) < E_1^\both \,,
  \qquad
  \both\in\{D,DN\}.
$$
Let $k \not\equiv 0$ be a periodic function of a period~$L>0$,
\ie, $\forall s\in\Real: \ k(s+L)=k(s)$,
and such that the hypothesis~$\Hi$ holds true for some~$d>0$.
Then the operator $H^\both$ is invariant with respect to the
transformation $s \mapsto s + j L$ for every $j \in \Int$,
which implies that there is no discrete eigenvalue in its spectrum,
\ie~$\sigma (H^\both) = \sigma_\mathrm{ess} (H^\both)$.
However, Theorems~\ref{Thm.Disc.Dirichlet} and~\ref{thm.DN}
state that
$
  \inf\sigma(H^\both) < E_1^\both.
$

According to a common belief, second order elliptic differential
operators with sufficiently regular
periodic coefficients should not have degenerate
bands in their spectra, or, in other words, their spectra should
be purely absolutely continuous (see~\cite{SS} and references therein).
An elegant rigorous proof of this fact for Dirichlet and Neumann
periodic waveguides was given by E.~Shargorodsky and A.~Sobolev in~\cite{SS}
(\cf~also~\cite{BDE} for thin Dirichlet tubes).
\end{exa}

The precedent Lemma~\ref{Lemma.Ess1} together with the following
one establish Theorem~\ref{Thm.Ess}.
\begin{lmm}\label{Lemma.Ess2}
If~$\di$ holds true, then
$
  \sigma_\mathrm{ess}(H^\both) \supseteq [E_1^\both,\infty) .
$
\end{lmm}
\begin{proof}
Let $n\in\Nat\setminus\{0\}$.
We shall construct a sequence $\{\psi_n^\both\}$ satisfying (i)--(iii)
of Lemma~\ref{Iftimie} with $\eta^\both:=\lambda^2+E_1^\both$
for all~$\lambda\in\Real$.
We start with the following family of functions
$$
  \hat{\psi}_n^\both(s,u)
  :=\varphi_n(s) \, \chi_1^\both(u) \,
  e^{i \lambda s} \,,
$$
where~$\chi_1^\both$ is the lowest transverse-mode
function~(\ref{TransEF}) if~$\both\in\{D,DN\}$,
or~(\ref{TransEF.N}) if~$\both=N$, respectively,
and $\varphi_n(s):=\varphi(n^{-1}s-n)$ with~$\varphi$ being
a non-zero $\Smooth^\infty$-smooth function 
with a compact support in $(-1,1)$.
Note that~$\supp\varphi_n \subset (n^2-n,n^2+n)$
and, consequently, the sequence~$\{\varphi_n\}$
is ``localized at~$+\infty$'' for large~$n$.
It is clear that~$\hat{\psi}_n^\both$ belongs to the form domain of~$H^\both$.
Since it is not normalized in~$\Hilbert$,
we introduce $\psi_n^\both:=\hat{\psi}_n^\both/\|\hat{\psi}_n^\both\|$.
Hereafter we shall use the equivalence
of the norms~$\|\cdot\|$ and~$\|\cdot\|_{\sii(\Omega_0)}$,
which follows by~(\ref{1<G<1}).
In particular, one has
\begin{equation}\label{NormEstimate}
  C_- \|\varphi_n\|_{\sii(\Real)}^2
  \leq
  \big\|\hat{\psi}_n^\both\big\|^2
  \leq
  C_+ \|\varphi_n\|_{\sii(\Real)}^2
\end{equation}
due to the normalization of~$\chi_1^\both$.

The point~(ii) of Lemma~\ref{Iftimie} requires that 
$(\phi,\psi_n^\both) \to 0$ as~$n\to\infty$ for all $\phi\in\Hilbert$.
Since~$\{\psi_n^\both\}$ is bounded in~$\Hilbert$,
it is enough to show the limit for all~$\phi\in\Smooth_0^\infty(\Omega_0)$,
a dense subset of~$\Hilbert$.
However, the latter follows at once because~$\phi$ and~$\psi_n^\both$
will have disjoint supports for~$n$ large enough.   

Hence, it remains to check
that \mbox{$\|(H^\both-\eta^\both)\psi_n\|_{-1} \to 0$}
as $n\to\infty$.
Employing the diagonal form~(\ref{metric}) of the metric tensor,
we can split the Hamiltonian~(\ref{Hamiltonian})
into a sum of two parts, $H^\both=H_1^\both+H_2^\both$,
where~$H_i^\both$, $i\in\{1,2\}$, corresponds to the term with $G^{ii}$
in~(\ref{Hamiltonian}).
This decomposition leads to the trivial bound
\begin{equation}\label{H<H1+H2}
  \big\|\big(H^\both-\eta^\both\big)\psi_n^\both\big\|_{-1}
  \leq
  \big\|\big(H_1^\both-\lambda^2\big)\psi_n^\both\big\|_{-1}
  +\big\|\big(H_2^\both-E_1^\both\big)\psi_n^\both\big\|_{-1} \,.
\end{equation}
We will show that the norms at the r.h.s. of this inequality
tends to zero as~$n\to\infty$ separately.
Denote
$
  \|f\|_{\infty,n}:=\sup\left\{|f(s,u)|\left|\,
  (s,u)\in\supp\varphi_n\times\trans\right.\right\}.
$

An explicit calculation using~(\ref{1<G<1})
and the fact that~$\chi_1^\both$
is an eigenfunction of~$-\Delta_\both^\trans$ corresponding
to the energy~$E_1^\both$ yields
\begin{equation}\label{easy}
  \left|
  \left(\phi,\big(H_2^\both-E_1^\both\big)\hat{\psi}_n^\both\right)
  \right|
  =\left|
  \big(\phi,k\,\hat{\psi}^\both_{n,2}\big)_{\!\sii(\strip_0)}
  \right|
  \leq
  C_-^{-1} \sqrt{E_1^\both} \
  \|k\|_{\infty,n}\,\|\phi\|\,\|\hat{\psi}_n^\both\|
\end{equation}
for all~$\phi\in\Dom Q^\both$.
Consequently, the second term at the r.h.s. of~(\ref{H<H1+H2})
goes to zero as~$n\to\infty$ by the assumption~$\di$.

A little more toilsome but still direct calculation yields
\begin{eqnarray*}
  \left(\phi,\big(H_1^\both-\lambda^2\big)\hat{\psi}_n^\both\right)
&=&
  \lambda^2 \left(\phi,
  \big(1-G^\demi\big)\,\hat{\psi}^\both_n\right)_{\!\sii(\strip_0)}
  \\
&& + \ \left(\phi_{,1},\big(G^\demi G^{11}-1\big)\,
  (\dot{\varphi}_n+i\lambda\varphi_n)\,\chi_1^\both e^{i\lambda s}
  \right)_{\!\sii(\strip_0)}
  \\
&& - \ \left(\phi,(\ddot{\varphi}_n+2i\lambda\dot{\varphi}_n)\,
  \chi_1^\both e^{i\lambda s}\right)_{\!\sii(\strip_0)}
\end{eqnarray*}
for all~$\phi\in\Dom Q^\both$.
Estimating all the terms at the r.h.s. of this equality
in the same way as in~(\ref{easy}),
it is enough to show that the following sequences
\begin{equation*}
  \big\|1-G^\demi\big\|_{\infty,n}, \quad
  \big\|G^\demi G^{11}-1\big\|_{\infty,n}, \quad
  \frac{\|\dot{\varphi}_n\|_{\sii(\Real)}}{\|\varphi_n\|_{\sii(\Real)}}
  \,, \quad
  \frac{\|\ddot{\varphi}_n\|_{\sii(\Real)}}{\|\varphi_n\|_{\sii(\Real)}}
  \,,
\end{equation*}
has the zero limit as~$n\to\infty$.
However, this is evident for the first and second ones
by virtue of~(\ref{metric}) and~$\di$,
while for the rest it follows by the definition of the sequence~$\{\varphi_n\}$.
\end{proof}

If the strip is asymptotically straight, \ie~$\di$,
then $\sigma(H^N)=[0,\infty)$ by Theorem~\ref{Thm.Ess}, (\ref{TransEV})
and non-negativity of~$H^N$; see also Theorem~\ref{thm.N}.
We conclude this section by proving the following result
about the spectral threshold of the operators~$H^D$ and~$H^{DN}$.
\begin{prp}\label{Prop.lowerbound}
Suppose~$\Hi$. If the strip obeys~$\di$, then
$$
  \inf\sigma(H^\both)>0
  \qquad\textrm{for}\quad\both\in\{D,DN\}.
$$
\end{prp}
\begin{proof}
We have~$H^\both \geq 0$ and~$E_1^\both>0$.
By virtue of Theorem~\ref{Thm.Ess}, it is enough to prove
that $0\not\in\sigma_\mathrm{p}(H^\both)$.
Assume that there exists $\psi\in\Dom H^\both$ such that $H^\both\psi=0$.
Then $\psi\in\Dom Q^\both$ and
$
  0=(\psi,H^\both\psi)=Q^\both[\psi]\equiv
  \int_{\strip_0} \overline{\psi}_{,i} G^{ij} \psi_{,j} \,d\strip
$
with~$(G_{ij})$ being a strictly positive definite matrix,
hence $\psi=0$ a.e.
\end{proof}
\noindent
Actually, stronger lower bounds to~$\inf\sigma(H^D)$
were derived in~\cite{AE,EFK}.

\section{Curvature-induced spectrum}\label{Sec.Dis}
Now we will be interested in the proofs concerning
the existence and properties of the spectrum of~$H^\both$
below the energy $E_1^\both$.
Since~$H^N$ is a non-negative operator and~$E_1^N=0$,
only the situations $\both\in\{D,DN\}$ are relevant here,
however, we do not exclude the Neumann case
from the preliminary considerations here
in order to establish a minor result
contained in Proposition~\ref{Prop.upperbound} below.

All the proofs of the following subsections
are based on the variational strategy of finding
a trial function~$\psi^\both$ from the form domain of $H^\both$
such that
\begin{equation}\label{Vfunctional}
  Q^\both_1[\psi^\both]
  \,:=\,
  Q^\both[\psi^\both]\,-\,E_1^{\both}\,\|\psi^\both\|^2\,<\,0 .
\end{equation}
We construct such a trial function by modifying the generalized
eigenfunction~(\ref{TransEF}) of energy~$E_1^\both$ for the straight strip.
This idea goes back to J.~Goldstone and R.~L.~Jaffe, \cite{GJ};
see also~\cite{DE,DEK2,DKriz2,K1}.

As a preliminary, let us express the form~(\ref{Vfunctional})
in the situation where the variables are separated
in the following way:
\begin{equation}\label{separation}
  \psi^\both(s,u) := \varphi(s)\,\chi^\both_1(u),
\end{equation}
where~$\chi_1^\both$ is the first transverse mode~(\ref{TransEF})
or~(\ref{TransEF.N})
and~$\varphi$ is a suitable function from $\Sobi(\Real)$.
In view of~(\ref{1<G<1}),
it is clear that~$\psi^\both$ belongs to $\Dom Q^\both$, given
by~(\ref{Dom.Dform}), (\ref{Nform}) or~(\ref{DNform}), respectively.
An explicit calculation yields
\begin{equation}\label{GJ-step0}
  Q_1^\both[\psi^\both ]
  =\left(\dot{\varphi},\langle G^{-\demi} \rangle_\both \,
  \dot{\varphi}\right)_{\sii(\Real)}
  + \mbox{$\demi$}
  \left[\chi_1^\both(d)^2-\chi_1^\both(0)^2\right]
  \left(\varphi,k\,\varphi\right)_{\sii(\Real)} \,,
\end{equation}
where~$\langle\cdot\rangle_\both$ denotes the expectation
with respect to $\chi_1^\both$, \ie\/
$$
  \langle f \rangle_\both :=
  \int_\trans f(\cdot,u) \chi_1^\both(u)^2 du
$$
with~$f\in\sinf(\Omega_0)$.
It is clear from~(\ref{TransEF}) and~(\ref{TransEF.N})
that the second term at the r.h.s. of~(\ref{GJ-step0})
is absent for~$\both\in\{D,N\}$,
while~$\chi_1^{DN}(d)=\sqrt{2/d}$ and~$\chi_1^{DN}(0)=0$.

\subsection{The existence}\label{Sec.Existence}
%
\paragraph{\emph{Proof of} Theorem~\ref{Thm.Disc.Dirichlet}
and Theorem~\ref{thm.DN}, \emph{condition} (a).}
%
We set
\begin{equation}\label{trial.function}
  \psi_n^\both(s,u):=\varphi(s;n)\,\chi_1^\both(u),
\end{equation}
where~$\varphi : \Real\times(0,\infty) \to [0,1]$
is supposed to satisfy:
\begin{itemize}
\item[(i)]
  $\forall n\in(0,\infty): \ \varphi(\cdot;n) \in \Sobi(\Real)$,
\item[(ii)]
  $\varphi(s;n) \xrightarrow[n \to \infty]{} 1$ \quad
  for a.e. $s \in \Real$,
\item[(iii)]
  $\|\varphi_{,1}(\cdot;n)\|_{\sii(\Real)}
  \xrightarrow[n \to \infty]{} 0$,
\end{itemize}
that is, $\varphi$ is a suitable mollifier of~$1$
(for an example of such a function, see~(\ref{mollifier}) below).
Substituting this trial function to~(\ref{GJ-step0}), we get
\begin{equation}\label{GJ-step1}
  Q_1^D[\psi_n]
  \ \xrightarrow[n \to \infty]{} \ 0 \,,
  \qquad
  Q_1^N[\psi_n]
  \ \xrightarrow[n \to \infty]{} \ 0 \,,
  \qquad
  Q_1^{DN}[\psi_n]
  \ \xrightarrow[n \to \infty]{} \
  \frac{\alpha}{d}
  \,,
\end{equation}
where~$\alpha$ is the total bending angle~(\ref{bending angle}).
The limits hold true by virtue of the required properties
of~$\varphi$, the fact that $\langle G^{-\demi} \rangle_\both$
are bounded functions and, in the case $\iota = DN$, also by
the dominated convergence theorem.
That is why we need to assume in addition that~$k$ is integrable
for~$\both=DN$.
Consequently, if~$\alpha$ is strictly negative,
then there exists a finite~$n_0>0$ such that $Q_1^{DN}[\psi_{n_0}]<0$
and the proof for $\iota = DN$ is finished in this case.

To obtain the result for~$\both=DN$ in the limit case~$\alpha=0$,
and for any Dirichlet strip,
we modify the function~$\psi_n^\both$,
in a curved part of the waveguide.
We define
\begin{equation}\label{trial.function.eps}
  \psi_{n,\varepsilon}^\both(s,u) :=
  \psi_n^\both(s,u)
  +\varepsilon\,\phi(s)\,\upsilon^\both(u)\,\chi_1^\both(u),
  \qquad
  \both\in\{D,DN\}
\end{equation}
where $\varepsilon\in\Real$,
$\phi\in\Sobi(\Real)$ is a real, non-negative, non-zero
function with compact support
contained in a bounded interval in~$\Real$ where~$k$
is not zero and does not change sign
(such an interval surely exists because~$k\not\equiv0$
and is continuous),
and $\upsilon^D(u):=-2u/d$ and $\upsilon^{DN}(u):=1$.
The family~$\{\psi_{n,\varepsilon}^\both\}$ is a subset of $\Dom Q^\both$
and we can write
\begin{equation}\label{GJ-step2}
  Q_1^\both[\psi_{n,\varepsilon}^\both]
  =Q_1^\both[\psi_n^\both]
  +2\varepsilon \,
  Q_1^\both(\phi\upsilon^\both\chi_1^\both,\psi_n^\both)
  +\varepsilon^2 \, Q_1^\both[\phi\upsilon^\both\chi_1^\both].
\end{equation}
The last term at the r.h.s. of~(\ref{GJ-step2})
does not depend on~$n$, while the first one
tends to zero as~$n\to\infty$ by~(\ref{GJ-step1}).
An explicit calculation of the central term gives
(\cf~(\ref{GJ-step0}) for~$\both=DN$)
$$
  Q_1^\both(\phi\upsilon^\both\chi_1^\both,\psi_n)
  = \big(\dot{\phi},\langle \upsilon^\both G^{-\demi} \rangle_\both \,,
  \dot{\varphi}_n\big)_{\sii(\Real)}
  + \mbox{$\frac{1}{d}$}
  \left(\phi,k\,\varphi_n\right)_{\sii(\Real)},
$$
where we have denoted~$\varphi_n:=\varphi(\cdot;n)$
and $\dot{\varphi}_n:=\varphi_{,1}(\cdot;n)$.
Using then the properties of the function~$\varphi$
together with the dominated convergence theorem
(notice that $\phi\,k \in \si(\Real)$), we have
\begin{equation}\label{GJ-step2.5}
  Q_1^\both[\psi_{n,\varepsilon}^\both]
  \xrightarrow[n\to\infty]{}
  \mbox{$\frac{2}{d}$}\,\varepsilon \left(\phi,k\right)_{\sii(\Real)}
  +\varepsilon^2 \, Q_1^\both[\phi\upsilon^\both\chi_1^\both].
\end{equation}
Since the integral~$(\phi,k)_{\sii(\Real)}$ is non-zero
by the construction of~$\phi$, 
we can take~$\varepsilon$ sufficiently small
and of an appropriate sign
so that the sum of the last two terms at the r.h.s. of~(\ref{GJ-step2.5})
is negative, and then choose~$n$ sufficiently large
so that $Q_1^\both[\psi_{n,\varepsilon}^\both]<0$.
\qed
\medskip

The intermediate results~(\ref{GJ-step1}) of the precedent proof
give the following upper bounds to the spectral threshold of~$H^\both$:
\begin{prp}\label{Prop.upperbound}
Suppose~$\Hi$. One has
\begin{itemize}
\item[\emph{(i)}]
$
  \inf\sigma(H^\both) \leq E_1^\both 
$
\quad for \ $\both\in\{D,N\};$
\item[\emph{(ii)}] 
$
  \inf\sigma\left(H^{DN}-\frac{k(s)}{d(1-uk(s))}\right) \leq E_1^{DN}
$
\quad provided \ $k\in\si(\Real)$.
\end{itemize}
\end{prp}
\noindent
Actually, in view of Theorem~\ref{Thm.Disc.Dirichlet},
a stronger result than~(i) holds for any Dirichlet strip.
The assertion~(i) for the Neumann case,
together with the fact that~$H^N$ is non-negative,
establishes the first claim of Theorem~\ref{thm.N}.
\begin{rem}[Condition~(a) of Theorem~\ref{thm.DN} \emph{vs}
the assumptions in~\cite{DKriz2}]\label{Rem.comparison}
The non-positivity of the total bending angle, \ie~$\alpha \leq 0$,
is a nice sufficient condition which guarantees
the existence of geometrically induced spectrum for~$H^{DN}$.
This was established already in~\cite{DKriz2}
under the additional hypothesis that
``$k$ is non-positive everywhere outside of some bounded interval''.
Since the latter is not assumed in this paper,
we extend significantly the class of admissible geometries.
Nevertheless, in order to justify the use of the dominated convergence theorem,
we need to assume that ``$k$ is integrable'' instead;
\cf~the condition~(a) of Theorem~\ref{thm.DN}.
Hence, a natural question is to ask whether the assumptions in~\cite{DKriz2}
may after all present an alternative criterion
which is not contained in our condition~(a).
The answer is negative due to the following (purely geometrical) result,
which can be easily shown using
the so-called ``Umlaufsatz",~\cite[Thm.~2.2.1]{Kli}:
\begin{lmm}\label{Lem.Overlap}
Let~$\curve$ be an infinite plane $\Smooth^2$-smooth curve of
bounded curvature~$k$. If there exists
a~compact~$\curve_c\subset\curve$ such that $\big|\int_{\curve_a}
k\big| > 2\pi $ for any compact~$\curve_a$ obeying $\curve_c
\subseteq \curve_a \subset \curve$, then any tubular neighbourhood
of~$\curve$ overlaps.
\end{lmm}
\noindent
That is, any reference curve satisfying the assumptions of~\cite{DKriz2}
but having a non-integrable curvature
leads to a violation of the basic hypothesis~$\Hi$
(which is assumed in~\cite{DKriz2} as well).
\end{rem}
%

\paragraph{\emph{Proof of} Theorem~\ref{thm.DN}, \emph{condition} (b).}
Let~$L>0$ be the period of~$k$, \ie, $\forall s\in\Real:\ k(s+L)=k(s)$.
We take the trial function of the form
\begin{equation*}
\psi_{n,\varepsilon}^{DN}(s,u)
:=
\varphi_n(s)\,\left( 1 + \varepsilon\,\phi(s) \right)\,
\chi_1^{DN}(u)\,,
\end{equation*}
\cf~(\ref{trial.function.eps}), where the
functions~$\varphi_n$ and~$\phi$ are defined as follows.
Let $\varphi_1 \in \Smooth_0^\infty(\Real)$ be a real function
with the support inside the interval~$(-L,2L)$ which
is equal to~$1$ on the period cell~$(0,L)$.
We set, for any $n\in\Nat\setminus\{0\}$,
$$
  \varphi_n(s) :=
  \begin{cases}
    \varphi_1(s)
    & \ \textrm{if} \quad s \in (-\infty,L), \\
    1
    & \ \textrm{if} \quad s \in [L,n L], \\
    \varphi_1\left(s-(n-1)L\right)
    & \ \textrm{if} \quad s \in \left(nL,+\infty\right) .
 \end{cases}
$$
Let~$\phi\in\Smooth^\infty(\Real)$ be non-negative, $L$-periodic,
and such that \mbox{$\supp\phi\!\upharpoonright\!(0,L)$} is contained
in an interval where~$k$ is not zero and does not change sign.
Then $(\phi,k)_{\sii((0,L))}\not=0$.
Finally, let~$\varepsilon\in\Real$ be chosen in such a way that
(\cf~(\ref{GJ-step2.5}))
\begin{align}
  A
  \ := & \ \left(\psi_{1,\varepsilon}^{DN},
  \big(H^{DN}-E_1^{DN}\big) \psi_{1,\varepsilon}^{DN}
  \right)_{\sii\left((0,L)\times\trans,d\strip\right)}
  \label{A} \\
  \ = & \ \mbox{$\frac{2}{d}$}\,\varepsilon \left(\phi,k\right)_{\sii((0,L))}
  +\varepsilon^2 \,
  \left(\phi\chi_1^{DN},
  \big(H^{DN}-E_1^{DN}\big)\phi\chi_1^{DN}
  \right)_{\sii\left((0,L)\times\trans,d\strip\right)} \nonumber
\end{align}
is negative.
By virtue of the definition of~$\varphi_1$ and the fact that
$\int_0^L k(s) ds = 0$ (\cf~Lemma~\ref{Lem.Overlap}),
it is clear that
$$
  Q_1^{DN}[\psi_{1,\varepsilon}^{DN}] = A+B,
$$
where~$B$ is defined as the integral at the first line of~(\ref{A}),
however, with the range of integration
being the set~$((-L,0)\cup(L,2L))\times\trans$.
Using the periodicity of the coefficients of~$H^{DN}$
together with the definition of~$\varphi_n$,
we continue by induction and arrive at the identity
$$
  \forall n\in\Nat\setminus\{0\}: \quad
  Q_1^{DN}[\psi_{n,\varepsilon}^{DN}] = nA+B,
$$
which becomes negative for~$n$ sufficiently large.
\qed
\begin{rem}[Integrability of~$k$]
If~$k \not\equiv 0$ is periodic, then the curvature is not integrable.
However, one has for every~$n \in \Nat$,
$\int_{-n L}^{n L} k(s)\,ds=0$ due to the periodicity
(\cf~Lemma~\ref{Lem.Overlap}).
This indicates that the requirement~$k\in\si(\Real)$
in the condition~(a) of Theorem~\ref{thm.DN}
may be rather a technical hypothesis.
\end{rem}
%

\paragraph{\emph{Proof of} Theorem~\ref{thm.DN}, \emph{condition} (c).}
We take the trial function~$\psi^{DN}$ of the
form~(\ref{separation}).
Since $k$ is continuous and $k_- \not\equiv 0$, there exists
an interval $J \subset \Real$, such that $k(s) < 0$ for all $s \in J$.
Choosing $\varphi \in \Sobi(\Real)$
such that $\supp\varphi \subseteq \overline J$
and substituting it to~(\ref{GJ-step0}),
obvious estimates yield
\begin{equation}\label{VarResultThmDis2}
Q_1^{DN}[\psi^{DN}]\,\leq\,\left\|\dot\varphi \right\|^2_{L^2(J)}\, +
\, \frac{1}{d}\int_J |\varphi(s)|^2\, k(s)\,d s.
\end{equation}
The second term at the r.h.s. of the last inequality
is obviously negative, while the first one does not depend on~$d$.
Hence for all~$d$ sufficiently small their sum is negative.
\qed

\paragraph{\emph{Proof of} Proposition~\ref{Prop.Number}.}
The claim is trivial for $n=0$.
Let us fix an integer $n \in \Nat\setminus\{0\}$.
We shall find a critical width $d_n$ such that for all $d < d_n$,
there are at least $n$ discrete eigenvalues in the spectrum
of~$H^{DN}$, counting multiplicity.
As in the proof of Theorem~\ref{thm.DN}, condition~(c),
let~$J \subset \Real$ be a bounded interval such that~$k(s) < 0$
for all $s \in J$.
We set $s_0 := \inf J$ and $s_j:= s_0 + j \, |J| /n$
for every $j \in \{1,\dots,n\}$.
Let~$\varphi_0$ be a non-zero function from $\Sobi(\Real)$ such that
$\supp\varphi_0 \subset (s_0, s_1)$.
We define for every $j \in \{1,\dots,n\}$ and $s\in\Real$,
\begin{align*}
  N_j^{-2} &:= \int_{s_{j-1}}^{s_j} \left| \varphi_0
  (s_0 + s - s_{j-1})\right|^2\, \langle G^\demi \rangle_{DN} (s) \, d s
  \,,\\
  \varphi_j(s) &:= N_j\,\varphi_0(s_0 + s - s_{j-1})
  \,.
\end{align*}
Putting
$\psi_j^{DN}(s,u) := \varphi_j(s) \chi_1^{DN}(u)$
for every $j \in \{1,\dots,n\}$, \cf~(\ref{separation}),
we get an ortho\-normal basis of a subspace of $\Dom Q^{DN}$.
Moreover, $Q^{DN}(\psi_j^{DN},\psi_\ell^{DN}) = 0$
whenever $j \not= \ell$
because $\varphi_j$ and $\varphi_\ell$ have disjoint supports.
Therefore, it follows by~\cite[Lemma~4.5.4]{Davies}
and Theorem~\ref{Thm.Ess} that a sufficient condition for $H^{DN}$
to have at least $n$~discrete eigenvalue is
$Q^{DN}[\psi_j^{DN}] < E_1^{DN}$, \ie~$Q^{DN}_1[\psi_j^{DN}] < 0$,
for every $j\in \{1,\dots n\}$.
However, according to~(\ref{VarResultThmDis2}),
\begin{equation*}
  Q_1^{DN} \left[\psi_j^{DN}\right] \ \leq \
  N_j^2 \left \| \dot\varphi_0 \right\|^2_{L^2(\Real)} +
  \frac{N_j^2}{d} \int_{s_{j-1}}^{s_j}
  \left| \varphi_0 (s_0 + s - s_{j-1})\right|^2 \, k(s) \, ds\,.
\end{equation*}
The r.h.s. of the last inequality is obviously negative for all
$j \in \{1,\dots,n\}$ provided that $d < d_n$ with
\begin{equation*}
  d_n := \min_{j \in \{1,\dots,n\}} \frac{1}{\left \| \dot\varphi_0
  \right\|^2_{L^2(\Real)}}\,
  \int_{s_0}^{s_1} | \varphi_0 (s)|^2 \,
  \left|k(s - s_0 + s_{j-1})\right| \, d s\,.
\end{equation*}
\qed
%

\subsection{The estimates on the spectral threshold}\label{Sec.Estimate}
%
Throughout this subsection,
we consider only $\both \in \{D, DN\}$.
Obviously,
\begin{equation}\label{minimax}
  \inf \sigma(H^\both)-E_1^\both
  =\inf_{\psi\in\Dom Q^\both} \frac{Q_1^\both[\psi]}{\|\psi\|^2}
  \leq \inf_{\psi \in T^\both} \frac{Q_1^\both[\psi]}
  {\|\psi\|^2} \,,
\end{equation}
where $T^\both$ is an arbitrary subset of $\Dom Q^\both$.
Our strategy will be to choose a suitable~$T^\iota$ and then
explicitly find the infimum of the quotient
at the r.h.s. of~(\ref{minimax}).

In Theorem~\ref{thm.estimate}, the curvature is supposed to have
a compact support contained in an interval of width~$2 s_0$;
without loss of generality we may assume that the reference
curve is parameterized in such a way that
$\supp k \subseteq [-s_0,s_0]$.

\paragraph{\emph{Proof of} Theorem~\ref{thm.estimate},
\emph{part} (i).}
%
Let~$\psi_{n,c}(s,u):=\varphi_c(s;n)\,\chi_1^{DN}(u)$ be the trial
function from the beginning of the proof of the condition~(a)
of Theorem~\ref{thm.DN} in Section~\ref{Sec.Existence}
with the mollifier~$\varphi_c(\cdot;n)$ given explicitly by
\begin{equation}\label{mollifier}
  \varphi_c(s;n) :=
  \begin{cases}
    1 & \textrm{if}\quad |s|\in [0,n), \\
    (c\,n-|s|)/\left((c-1)\,n\right) & \textrm{if}\quad |s|\in[n,c n), \\
    0 & \textrm{if}\quad |s|\in [c n,\infty),
  \end{cases}
  \qquad c>1.
\end{equation}
We set $T^{DN} := \left\{ \psi_{n,c}\,|\,n \geq s_0\,,\, c > 1 \right\}$.
An easy calculation yields
\begin{align*}
  Q^{DN}_1[\psi_{n,c}]=\frac{2}{(c-1)\,n}+\frac{\alpha}{d} \,,
  \qquad
  \|\psi_{n,c}\|^2
  =\frac{2}{3}\,(c+2) \,n - \alpha\,\langle u \rangle \,,
\end{align*}
where
\begin{equation*}
  \langle u \rangle
  := \int_\trans u\,\chi_1^{DN}(u)^2\,du
  =d\,\left(\frac{1}{2}+\frac{2}{\pi^2}\right).
\end{equation*}
Hence, denoting by~$f(n,c)$ the quotient
at the r.h.s. of~(\ref{minimax}), we have
\begin{equation}
  f(n,c)=\frac{\frac{2}{c-1}+\frac{\alpha}{d}n}
  {\frac{2}{3} (c+2)\,n^2 - \alpha\,\langle u \rangle\,n}
  \,.
\end{equation}
Now we shall seek the infimum of the continuous function $f$
in the region $[s_0,\infty)\times (1,\infty)$;
the result establishes the bound from Theorem~\ref{thm.estimate}.

One can directly check that there is no local minimum of
the function~$f$ in the interior of its domain, \ie~for every point
$(n,c) \in (s_0,\infty) \times (1,\infty)$,
$f_{,1}(n,c) \neq 0$ or $f_{,2}(n,c) \neq 0$.
Thus the problem reduces to the study of the behaviour of~$f$
on the boundary set $\{s_0\}\times (1,\infty)$
and its limits as $n \to \infty$, $c \to 1$ and $c \to \infty$,
respectively.
The function~$f(s_0,\cdot)$ reaches its (negative) local minimum
\begin{equation}\label{minimal.value}
  f\left(s_0,c_+\right)
  =\frac{-3\,\alpha^2/d^2}
  {4\,\left(1+\sqrt{1-\frac{3}{2} \alpha\,s_0 / d
  +\frac{3}{4}\alpha^2\langle u \rangle/d}\right)^2} \,
\end{equation}
for
\begin{equation}\label{n+}
  c_+ := - \frac{2 d}{\alpha\,s_0} + 1 - \frac{d}
  {\alpha\,s_0}\,\sqrt{-6\frac{\alpha\,s_0}{d} + 4 + 3
  \frac{\alpha^2\, \langle u\rangle}{d}}
   \,.
\end{equation}
Using the estimate
$
  f(n,c) \geq \alpha/(2 d n),
$
we obtain
\begin{equation*}
\liminf_{n\to\infty} f(n,c) \geq 0
\end{equation*}
uniformly in~$c$.
Hence, there exists a (finite)~$n_0 > s_0$ such that for
every~$n > n_0$ holds true $f(n,c) \geq f(s_0,c_+)$
(recall that $f(s_0,c_+) < 0$, \cf~\eqref{minimal.value})
uniformly in~$c$.
Therefore since we seek the infimum of~$f$
we can consider only~$n \in [s_0,n_0]$ in the rest of the
proof.
However, for those values of~$n$ we have
\begin{equation}\label{Thm.Estimate.Limit}
f(n,c) \geq \frac{6}{(c-1)\left( 2(c+2) n_0^2 - 3 \alpha
\langle u \rangle n_0 \right)} +\frac{3\alpha / d}{2 (c+2) n
-3 \alpha \langle u \rangle}
\end{equation}
and since
\begin{align*}
\lim_{c \to 1}\frac{6}{(c-1)\left( 2(c+2) n_0^2- 3 \alpha
\langle u \rangle n_0 \right)} = \infty\,,\quad
\left|\lim_{c \to 1}\frac{3\alpha / d}{2 (c+2) n - 3\alpha \langle u
\rangle} \right| < \frac{1}{d \langle u \rangle} \,,
\end{align*}
we obtain
$$
  \lim_{c\to 1} f(n,c) = \infty
$$
uniformly in~$n$.
Finally,
\begin{align*}
\lim_{c \to \infty}\frac{6}{(c-1)\left( 2(c+2) n_0^2 - 3 \alpha
\langle u \rangle n_0 \right)} &= 0\,,\\
\lim_{c \to \infty}\frac{3\alpha / d}{2 (c+2) n -3 \alpha \langle u
\rangle} \geq \frac{\alpha}{d n}\,\lim_{c \to \infty}\frac{3} {2
(c+2)} &=0
\end{align*}
because $n \mapsto \alpha / (d n)$ is bounded on~$[s_0,n_0]$;
hence, in view of~(\ref{Thm.Estimate.Limit}),
\begin{equation*}
\liminf_{c\to\infty} f(n,c) \geq 0
\end{equation*}
uniformly in~$n$.
Since the infimum of~$f$ should be negative,
we infer from the above results that
$$
  \inf_{(n,c)\in [s_0,\infty)\times(1,\infty)} f(n,c)
  = \inf_{c\in(1,\infty)} f(s_0,c)
  = f\left(s_0,c_+\right) ,
$$
where $f(s_0,c_+)<0$ given by~(\ref{minimal.value})
provides an upper bound on the r.h.s.
of~(\ref{minimax}) for the case $\both = DN$.
\qed
%
\paragraph{\emph{Proof of} Theorem~\ref{thm.estimate},
\emph{part} (ii).}
In the Dirichlet case, we use the mollifier~(\ref{mollifier})
with the fixed $n=s_0$ for the construction of the functions
from $T^D$.
We set for any $c_1,\,c_2 > 1$
and $\varepsilon \in \Real$,
\begin{equation}\label{Thm.Estimate.Trial}
  \psi_{c_1,c_2,\varepsilon}(s,u) := \varphi_{c_1}(s;s_0)\,
  \chi_1^D(u) + \varepsilon\,\varphi_{c_2}(s;s_0)\, \chi_2^D(u)
\end{equation}
and
$
  T^D := \{ \psi_{c_1,c_2,\varepsilon}\,|\,c_1,\,c_2 >1\,,\,
  \varepsilon \in \Real\}
$.
Easy explicit calculations give
\begin{align*}
Q_1^D[\psi_{c_1,c_2,\varepsilon}] &=
\frac{\pi^2}{d}\,\left( h(c_1) + \frac{16}{3 \pi^2} \alpha
\varepsilon + \varepsilon^2 (2 g(c_2) + h (c_2)) \right)\,,\\
\\
\left\|\psi_{c_1,c_2,\varepsilon} \right\|^2 &=
\frac{2d}{3} \left( g(c_1) + \frac{16}{3 \pi^2} \alpha
\varepsilon + \varepsilon^2 g(c_2)\right)\,,
\end{align*}
where
\begin{equation*}
h(c):=
\frac{2}{\pi^2}\,\frac{d}{s_0}\,\frac{1}{c-1}\,,
\qquad
g(c):=
\frac{s_0}{d}(c+2)-\frac{3}{4} \alpha\,.
\end{equation*}
Thus, the quotient at the r.h.s. of~(\ref{minimax}) can be written
as
\begin{equation} \label{tilde.f}
  \tilde f(c_1, c_2, \varepsilon) :=
  \frac{3 \pi^2}{2 d^2}\, \frac{h(c_1) + \frac{16}{3 \pi^2} \alpha
  \varepsilon + \varepsilon^2 (2 g(c_2) + h (c_2))}
  { g(c_1) + \frac{16}{3 \pi^2} \alpha
  \varepsilon + \varepsilon^2 g(c_2)}\,.
\end{equation}
Clearly, $\tilde f$ is a continuous function of the three variables
defined in the region $(1,\infty)^2 \times \Real$
(the denominator is positive since
it is the squared norm of a nonzero function)
and one could look for its infimum.
However, from the technical point of view,
it seems to be a rather complicated task
and that is why we make first the following simplification.

We start by verifying that the infimum of~$\tilde{f}$ is negative,
\ie, $\psi_{c_1,c_2,\varepsilon}$ is an admissible
trial function to estimate $\inf(H^D)-E_1^D<0$,
\cf~Theorem~\ref{Thm.Disc.Dirichlet}.
Obviously, $h(c) > 0$ for any $c \in (1,\infty)$.
Using the definition of~$\alpha$, the assumption~$\Hi$
and obvious estimates, we check that the same holds true for~$g$:
\begin{equation}\label{Thm.Estimate.g}
g(c) > 3 \left( \frac{s_0}{d} - \frac{\alpha}{4}\right) 
> 3 \left(\frac{s_0}{d} - \frac{1}{2} s_0\|k_+\|_\infty
\right) > \frac{3}{2}\,\frac{s_0}{d}\,.
\end{equation}
Hence, the only term in the numerator of $\tilde f$ which can
attain negative values is the term linear in $\varepsilon$.
However, for any given $c_2 > 0$, there exists
$\varepsilon \in \Real$ of such a sign
that $\alpha\,\varepsilon < 0$
and with a sufficiently small absolute value
so that the negative term
linear in $\varepsilon$ dominates over the quadratic one.
Then we can find $c_1$ large enough to make the numerator of the
r.h.s. of~(\ref{tilde.f}) negative.
Recalling that the denominator is positive, we can restrict
ourselves to those values of the triple
$(c_1,c_2, \varepsilon)$, for which
$\tilde f(c_1,c_2,\varepsilon) <0$;
let us denote $\mathcal{N} := \{ (c_1,c_2,\varepsilon)\in
(1,\infty)^2 \times \Real\,|\,\tilde f(c_1,c_2,\varepsilon) <
0\}$.
Setting for any $(c_1,c_2,\varepsilon) \in \mathcal{N}$,
\begin{equation}\label{f(c1,c2,epsilon)}
f(c_1,c_2,\varepsilon) := \frac{3 \pi^2}{2 d^2}\, \frac{h(c_1) +
\frac{16}{3 \pi^2} \alpha \varepsilon + \varepsilon^2
(2 g(c_2) + h (c_2))}
{ g(c_1)}\,,
\end{equation}
we arrive easily at the inequality
$
  \tilde f(c_1,c_2,\varepsilon)
  \leq f(c_1,c_2,\varepsilon)
$,
because the (positive) denominator in (\ref{tilde.f})
is bounded from above by $g(c_1)$ due to the above considerations.
Consequently,
\begin{equation}\label{minimax.D}
  \inf \sigma(H^D)-E_1^D \leq
  \inf_{(c_1,c_2,\varepsilon) \in \mathcal{N}} f
  (c_1,c_2,\varepsilon).
\end{equation}

Calculating the partial derivatives of~$f$,
it is straightforward to see that the system of equations
$f_{,i} = 0$, $i = 1,2,3$, can be cast into the following form:
\begin{align*}
\frac{s_0}{d}\, A(c_2, \varepsilon)\,(c_1-1)^2 + \frac{4}{\pi^2}\,
(c_1-1) + \frac{6}{\pi^2}\,\frac{d}{s_0}\,\left( \frac{s_0}{d} -
\frac{\alpha}{4} \right) &= 0\,,\\
(c_2 - 1)^2 - \left (\frac{d}{\pi s_0} \right)^2 &= 0\,,\\
\varepsilon + \frac{8 \alpha}{3 \pi^2}\, \frac{1} {h(c_2) + 2
g(c_2)} &=0\,,
\end{align*}
respectively, where, for any $(c_1,c_2,\varepsilon) \in\mathcal{N}$,
\begin{equation*}
A(c_2, \varepsilon) :=
\frac{16}{3 \pi^2} \alpha \varepsilon + \varepsilon^2
(2 g(c_2) + h (c_2)) < 0.
\end{equation*}
From the second equation we can immediately express $c_2$;
of course, we choose that root ${c_2}_+$ which is greater than~$1$.
Substituting  ${c_2}_+$ to the third equation of our system,
we obtain the root $\varepsilon_0$
(notice that really $\alpha\,\varepsilon_0 < 0$).
Finally, putting ${c_2}_+$ and $\varepsilon_0$ to the first equation,
we choose that root ${c_1}_+$ which is greater than~$1$.
A tedious but straightforward calculation yields
\begin{equation*}
  f({c_1}_+,{c_2}_+,\varepsilon_0) =
  - \frac{3 \pi^4}{4 d^2}\, \frac{A({c_2}_+,\varepsilon_0)^2} {\left(
  1 + \sqrt{ 1 - \frac{3}{2}A({c_2}_+,\varepsilon_0)\pi^2
  \left(\frac{s_0}{d}-\frac{\alpha}{4} \right)} \right)^2}
\end{equation*}
with
$$
  A({c_2}_+,\varepsilon_0)
  = -\frac{32\alpha^2}{9\pi^4}\frac{1}{\frac{2}{\pi}
  +3(\frac{s_0}{d}-\frac{\alpha}{4})}\,.
$$
(Recall that $\frac{s_0}{d} - \frac{\alpha}{4} >0$,
so the square root in the first formula is well defined in~$\Real$.)
Hence really $({c_1}_+,{c_2}_+,\varepsilon_0) \in \mathcal{N}$.
Moreover, one can check that the matrix of second derivatives of~$f$
is in the point $({c_1}_+,{c_2}_+,\varepsilon_0)$
diagonal with all positive elements,
that is, the function $f$ reaches its local
minimum in that point.

To see that it is the global minimum too, we study  the
behaviour of the limits of~$f$ as $c_i \to 1,\infty$,
$i\in\{1,2\}$ and $\varepsilon \to \pm\infty$.
We restrict ourselves to that cases, where the limit is reached by
negative values of~$f$; the rest of the ``boundary" of the
set~$\mathcal{N}$ consists of those triples $( \tilde c_1,
\tilde c_2, \tilde \varepsilon )$,
for which $f(\tilde c_1,\tilde c_2, \tilde \varepsilon)\,=\,0$,
that is,
$
  f(\tilde c_1,
  \tilde c_2, \tilde\varepsilon)\,>\,f(c_{1_+},c_{2_+},
  \varepsilon_0)
$.
Since~(\ref{Thm.Estimate.g}) gives
\begin{equation}\label{Thm.Estimate.g2}
  g(c) > \frac{3\,|\alpha|}{4}\,,
\end{equation}
we obtain
\begin{equation*}
  f(c_1,c_2,\varepsilon) > \frac{3\,\pi^2}{2\,d^2}\, \frac{
  \frac{16}{3\,\pi^2}\alpha\,\varepsilon + \frac{3\,
  |\alpha|}{2}\varepsilon^2}{g(c_1)}
\end{equation*}
and the condition~$f(c_1,c_2,\varepsilon) < 0$ yields
\begin{equation} \label{Thm.Estimate.epsilon}
|\varepsilon| < \frac{32}{9\,\pi^2}\,.
\end{equation}
Hence we do not study the limits as $\varepsilon \to
\pm \infty$ and we may assume in the following that $\varepsilon$
is bounded.
Using~(\ref{Thm.Estimate.g2}) in the denominator of~(\ref{f(c1,c2,epsilon)}),
neglecting~$h(c_1)$
and minimizing the remaining polynomial in~$\varepsilon$
in the numerator of~(\ref{f(c1,c2,epsilon)}),
we arrive at the lower bound
\begin{equation*}
  f(c_1,c_2,\varepsilon)
  > -\frac{128}{9\,\pi^2\,d^2}\,\frac{|\alpha|}{h(c_2) + 2\,g(c_2)}
\end{equation*}
for any $c_2\,\in\,(1,\infty)$.
Thus
\begin{equation*}
  \liminf_{c_2 \to \infty} f(c_1,c_2,\varepsilon) \geq 0\,,
  \qquad
  \liminf_{c_2 \to 1} f(c_1,c_2,\varepsilon) \geq 0
\end{equation*}
uniformly in $c_1$ and $\varepsilon$.
Finally, using~(\ref{Thm.Estimate.epsilon}) we can see that
\begin{equation*}
f(c_1,c_2,\varepsilon) > \frac{3\,\pi^2}{2\,d^2}\, \frac{h(c_1)}
{g(c_1)} - \frac{256\,|\alpha|}{9\,\pi^2\,d^2}\,\frac{1}{g(c_1)}
\end{equation*}
and therefore
\begin{equation*}
\liminf_{c_1\to\infty} f(c_1,c_2,\varepsilon) \geq 0\,,
\qquad
\lim_{c_1\to 1} f(c_1,c_2,\varepsilon) = \infty
\end{equation*}
uniformly in $c_2$ and $\varepsilon$.
Summing up the considerations, we conclude that
$f(c_{1_+},c_{2_+},\varepsilon_0)$
is the global minimum and
the claim~(ii) of Theorem~\ref{thm.estimate}
then follows from~(\ref{minimax.D}).
\qed
\begin{rem}[Mildly curved strips]\label{Rem.mildly} 
Let us compare our estimate~(ii) of Theorem~\ref{thm.estimate}
with the exact ground-state eigenvalue asymptotics
derived in~\cite[Thm.~4.1]{DE} for \emph{mildly} curved Dirichlet strips
by the Birman-Schwinger perturbation technique.
We consider families of generating curves $\curve_\beta$
characterized by the curvature $k_\beta(s) := \beta\,k(s)$,
where~$k$ is a fixed curvature function
and $\beta > 0$ is a small parameter.
Since $\alpha_\beta :=\int_\Real k_\beta(s)ds = \beta\,\alpha$,
we see that~$\beta$ controls the total bending of the strip, too.
The result of~\cite{DE} can be written as
\begin{equation*}
  \inf(H^{D}) = E_1^D\,-\,C(d,k)^2\,
  \beta^4 + \mathcal{O}(\beta^5)\,,
\end{equation*}
where~$C(d,k)$ is a positive constant depending only on the fixed width~$d$
and (integrals of)~$k$, while our estimate~(ii) yields
\begin{equation*}
  \inf(H^{D}) \leq E_1^D
  \, - \,
  C^D(s_0,d,0)^2 \, \alpha^4 \, \beta^4
  + \mathcal{O}(\beta^5)\,.
\end{equation*}
Hence we observe the same dependence of the leading terms
on the perturbation parameter~$\beta$.
Let us quantitatively compare the actual gap-width asymptotic
given by~$C(d,k)^2$ with our estimate $C^D(s_0,d,0)^2 \alpha^4$.
Since~$C(d,k)$ has rather a complicated structure,
we restrict ourselves to small values of the width~$d$ when
\begin{equation}\label{C(d,k)}
  C(d,k) =
  \frac{1}{8}\, \|k\|^2_{\sii(\Real)} +
  \mathcal{O}(d^2) \,.
\end{equation}
We have $C^D(s_0,d,0)=8/(9\sqrt{3}\pi^2 s_0)+\mathcal{O}(d)$.
Since $\alpha^2 \leq 2 s_0 \|k\|^2_{\sii(\Real)}$
by the Schwarz inequality, we see that
$$
  \frac{C^D(s_0,0,0)\,\alpha^2}{C(0,k)}
  \leq \frac{128}{9\sqrt{3}\,\pi^2}
  \approx 0.83 \,.
$$

As for the mixed Dirichlet-Neumann case,
our estimate~(i) of Theorem~\ref{thm.estimate} leads to
\begin{equation*}
  \inf(H^{DN}) \leq E_1^{DN}
  - \frac{3\,\alpha^2}{8\,d^2} \, \beta^2 +\mathcal{O}(\beta^3)
\end{equation*}
and we observe that the leading term is proportional
to the second power of~$\beta$ now.
In particular, it is much greater than the leading term
in the identical mildly curved strip
with the pure Dirichlet boundary condition.
Unfortunately, no exact asymptotics are known
for $\inf\sigma(H^{DN})$,
so we cannot perform any comparison in this case.
\end{rem}
\begin{rem}[Thin strips]\label{Rem.Thin}
Another natural perturbation parameter is the strip width~$d$.
Calculating the asymptotic expansions with respect to~$d$
of the constants $C^\both(s_0,d,\alpha)$
from our Theorem~\ref{thm.estimate},
we arrive at
\begin{align*}
  E_1^{DN} - \inf(H^{DN}) &\geq
  - \frac{\alpha}{2\,s_0\,d} + \mathcal{O} (d^{-{\demi}})\,,\\
  E_1^{D} - \inf(H^{D})&\geq
  \frac{2^8\,\alpha^4}{3^5\,\pi^4\,s_0^2\,\left(1+\sqrt{1+
  \left(\frac{4\alpha}{3\pi}\right)^2} \right)^2}
  +\mathcal{O}(d).
\end{align*}
Again, we observe qualitatively different behaviour
of our estimates with respect to the perturbation parameter.

In particular, the leading term in our lower estimate
of the gap between the essential spectrum threshold
and the lowest Dirichlet eigenvalue
is independent of the strip width.
This is in accordance with the perturbation expansion
of the ground-state eigenvalue derived in~\cite[Thm.~5.1]{DE}:
\begin{equation*}
  E_1^{D} - \inf(H^{D}) = - \lambda(k) + \mathcal{O}(d) \,.
\end{equation*}
Here~$\lambda(k)$ denotes the first (negative) eigenvalue
of the one-dimensional Schr\"{o}\-dinger operator
$
  l:=-\Delta\,-\,$\mbox{$\frac{1}{4}$}$\,k^2
$
on $\sii(\Real)$ with $\Dom l:= \Sobii(\Real)$,
which is naturally associated with the problem
and reflects the geometry of~$\curve$ only.
(We remark that, under our assumptions,
the operator~$l$ has always a negative eigenvalue,
\cf~\cite[Thm.~XIII.11]{RS4}.)

The leading term in the Dirichlet-Neumann estimate tends
to~$+\infty$ as $d \to 0$
(notice, however, that this fact does not conflict with anything
because~$E_1^{DN}=\mathcal{O}(d^{-2})$).
That is, we again observe the effect of stronger binding
of the particle in the case when a Dirichlet boundary curve
of the strip is replaced by the Neumann one.
A similar asymptotic estimate can be also deduced directly
from the crude bound~(ii) of Proposition~\ref{Prop.upperbound}.
Since no perturbation expansion with respect to~$d$ for the
lowest eigenvalue in the Dirichlet-Neumann case is known yet,
we cannot compare our estimate with exact asymptotics.
\end{rem}
%

\section{Conclusions}\label{Sec.Conclusions}
Motivated by the theory of curved quantum waveguides,
we were interested in spectral properties of the Laplace operator
in a strip built over an infinite planar curve, see Figure~\ref{Figure1},
subject to three different types of boundary conditions
(Dirichlet, Neumann or a combination of these ones, respectively).
We localized the essential spectrum as a set under a very
natural and weak condition about vanishing of curvature at infinity
only, \cf~Theorem~\ref{Thm.Ess}.
We stress that no condition about the decay of derivatives
of the curvature was required throughout this paper
(the derivatives may not even exist because the reference curve
is supposed to be~$\Smooth^2$-smooth only).
Then we were interested in the geometrically induced spectrum,
\ie\ the spectrum below the spectral threshold of
the corresponding straight strip;
we made a survey of known results and established new ones,
\cf~Theorems~\ref{thm.N}--\ref{thm.DN}.
Here the most important progress was achieved in the case
of combined Dirichlet-Neumann boundary conditions
where we generalized the only one known result of~\cite{DKriz2}
and established two new sufficient conditions which guaranteed
the existence of geometrically induced spectrum,
\cf~Theorem~\ref{thm.DN}.
We recall that the geometrically induced spectrum
consists of discrete eigenvalues only
whenever the above asymptotic behaviour of curvature holds true.
Finally, we established two upper bounds to the spectral threshold
in a situation when the geometrically
induced spectrum is present, \cf~Theorem~\ref{thm.estimate}.
These estimates are new in the theory of curved quantum waveguides
and their remarkable behaviour in the limit of mild curvature
or small width of the strip was discussed,
\cf~Remarks~\ref{Rem.mildly} and~\ref{Rem.Thin}.
Summing up briefly the main contribution of the paper,
we gave answers to the two questions formulated in
Section~\ref{Sec.Scope}.

Let us now mention some directions in which the above mentioned
results could be strengthened or extended.

In Theorem~\ref{Thm.Ess},
we succeeded to localize the essential spectrum as a set,
however, an open problem is to examine its nature.
Here a particularly interesting question is whether
the curved geometry may produce a singular continuous spectrum.
In the Dirichlet case, this problem was analysed quite recently in~\cite{KT}
by means of the Mourre theory.

Theorem~\ref{Thm.Disc.Dirichlet} concerning the existence
of geometrically induced spectrum in Dirichlet strips
is optimal in the sense that no better result can be achieved
without violating the basic hypothesis~$\Hi$.
One is of course tempted to ask which more general regions
(than the curved asymptotically straight strips)
still possess a non-trivial discrete spectrum.
For instance, it is easy to see that the existence result does not change
if the boundary of the strip is deformed locally and in such a way
that the resulting deformed region lies in the exterior of the strip,
\cf~\cite{RB}, however, more complicated deformations of the boundary
represent a difficult problem even in the straight case \cite{BGRS,BEGK}.
In this context, it is worth to recall that the existence
of discrete spectrum in \textsf{V}-shaped waveguides
was demonstrated in~\cite{ESS1,ABGM,CLM2} (the computed bound-state energy
has been verified experimentally in a flat electromagnetic
waveguide in~\cite{CLM1}).

The Neumann case is trivial from the point of view
of the existence of discrete spectrum
in asymptotically straight strips, \cf~Theorem~\ref{thm.N}.
As for the Dirichlet-Neumann strip,
while our Theorem~\ref{thm.DN} covers various wide classes of geometries
for which the geometrically induced spectrum exists,
it does not represent an ultimate result.
For instance, it remains to be clarified whether
one can include also some thick strips with a positive total bending angle.
Another open question concerning the strips
with combined boundary condition
is the study of the behaviour of eigenvalues in mildly curved,
respectively thin, strips, \cf~Remarks~\ref{Rem.mildly}
and~\ref{Rem.Thin}.

The upper bounds on the spectral threshold we presented
in Theorem~\ref{thm.estimate} can be surely improved.
First of all, one should include the situations when
the total bending angle is equal to zero and/or
the strip is curved globally.

As we have already mentioned in Introduction,
the Dirichlet Laplacian in the curved strip represents
a reasonable model for a quantum Hamiltonian
of a particle restricted to move in a strip-like nanostructure.
Assuming that the boundary is sufficiently regular,
to impose the Dirichlet boundary conditions means
to require the vanishing of wavefunctions, however,
as pointed out in~\cite{FTsutsui},
this may be in general too restrictive and one should rather require
the vanishing of the probability current only.
The latter leads in our case to a general boundary condition
of the type
$$
  a_0\psi(\cdot,0)+b_0\psi_{,2}(\cdot,0)=0 \,,
  \qquad
  a_d\psi(\cdot,d)+b_d\psi_{,2}(\cdot,d)=0 \,,
$$
where~$\psi\in\Hilbert$ denotes the wavefunction and $
  (a_0,a_d),(b_0,b_d) \in \Real^2\setminus\{(0,0)\}
$.
However, at least from the mathematical point of view,
it would be interesting to examine the influence
of the choice of particular boundary conditions
on the spectral properties of the Hamiltonian.
Finally, it would be also possible to let the coefficients
$a_0,a_d,b_0,b_d$ depend on the longitudinal variable~$s$.
 
Other obvious extensions are to consider the Laplacian
in tubular neighbourhoods of non-compact submanifolds
of general Riemannian manifolds.
Here the spectral problem was studied only for Dirichlet
tubes in~$\Real^3$ \cite{GJ,DE},
Dirichlet layers in~$\Real^3$ \cite{DEK2,CEK}
or more generally in~$\Real^n$ \cite{LL1},
and strips in two-dimensional manifolds~\cite{K1};
more general boundary conditions and other higher-dimensional
generalizations are still missing.

A long-standing open problem in the theory of quantum waveguides
is the question whether the geometrically discrete spectrum
in curved asymptotically straight Dirichlet strips
will ``survive" a strong homogeneous magnetic field.
In this context, let us mention the very recent work~\cite{MK-Kov}
(\cf~also \cite{B-MK-Kov}),
where it is shown actually that this is not the case for mildly curved strips
if an appropriate compactly supported magnetic field is added.

\section*{Acknowledgements}
\addcontentsline{toc}{section}{Acknowledgements}
One of the authors (D.K.)
would like to thank V.~Iftimie for pointing out to him
the general characterization of essential spectrum (Lemma~\ref{Iftimie})
which gives a better result than our original proof based
on the Weyl criterion; discussions on this subject with I.~Beltita
are also acknowledged.
We are very grateful to J.~Dittrich, P.~Duclos,
P.~Exner and P.~Freitas~for useful remarks and suggestions.
The first author was partially supported by FCT/POCTI/FEDER, Portugal,
and the AS\,CR project K1010104.
%
%

\begin{thebibliography}{10}

\bibitem{AE}
M.~S. Ashbaugh and P.~Exner, \emph{Lower bounds to bound state energies in bent
  tubes}, Phys. Lett. A~ \textbf{150} (1990), no.~3,4, 183--186.

\bibitem{APV}
A.~Aslanyan, L.~Parnovski, and D.~Vassiliev, \emph{Complex resonances in
  acoustic waveguides}, Q.~Jl~Mech. Appl. Math. \textbf{53} (2000), 429--447.

\bibitem{ABGM}
Y.~Avishai, D.~Bessis, B.~G. Giraud, and G.~Mantica, \emph{Quantum bound states
  in open geometries}, Phys.~Rev. B~ \textbf{44} (1991), 8028--8034.

\bibitem{BDE}
F.~Bentosela, P.~Duclos, and P.~Exner, \emph{Absolute continuity in periodic
  thin tubes and strongly coupled leaky wires}, Lett. Math. Phys. \textbf{65}
  (2003), 75--82.

\bibitem{B-MK-Kov}
D.~Borisov, T.~Ekholm, and H.~Kova{\v{r}}{\'\i}k, \emph{Spectrum of the
  magnetic {S}chr{\"o}dinger operator in a waveguide with combined boundary
  conditions}, math-ph/0405034.

\bibitem{BE}
D.~Borisov and P.~Exner, \emph{Exponential splitting of bound states in a
  waveguide with a pair of distant windows}, J.~Phys.~A \textbf{37} (2004),
  3411--3428.

\bibitem{BEG}
D.~Borisov, P.~Exner, and R.~Gadyl'shin, \emph{Geometric coupling thresholds in
  a two-dimensional strip}, J.~Math. Phys. \textbf{43} (2002), 6265--6278.

\bibitem{BEGK}
D.~Borisov, P.~Exner, R.~Gadyl'shin, and D.~Krej\v{c}i\v{r}\'{\i}k, \emph{Bound
  states in weakly deformed strips and layers}, Ann.~H.~Poincar{\'e} \textbf{2}
  (2002), 553--572.

\bibitem{BGRS}
W.~Bulla, F.~Gesztesy, W.~Renger, and B.~Simon, \emph{Weakly coupled bound
  states in quantum waveguides}, Proc. Amer. Math. Soc. \textbf{125} (1997),
  1487--1495.

\bibitem{CLM1}
J.~P. Carini, J.~T. Londergan, K.~Mullen, and D.~P. Murdock, \emph{Bound states
  and resonances in waveguides and quantum wires}, Phys.~Rev. B~ \textbf{46}
  (1992), 15538--15541.

\bibitem{CLM2}
\bysame, \emph{Multiple bound states in sharply bent waveguides}, Phys.~Rev. B~
  \textbf{48} (1993), 4503--4515.

\bibitem{CEK}
G.~Carron, P.~Exner, and D.~Krej\v{c}i\v{r}\'{\i}k, \emph{Topologically
  nontrivial quantum layers}, J.~Math.\ Phys. \textbf{45} (2004), 774--784.

\bibitem{Davies}
E.~B. Davies, \emph{Spectral theory and differential operators}, Camb. Univ
  Press, Cambridge, 1995.

\bibitem{DP}
E.~B. Davies and L.~Parnovski, \emph{Trapped modes in acoustic waveguides},
  Q.~Jl Mech. Appl. Math. \textbf{51} (1998), 477--492.

\bibitem{DDI}
Y.~Dermenjian, M.~Durand, and V.~Iftimie, \emph{Spectral analysis of an
  acoustic multistratified perturbed cylinder}, Commun. in Partial Differential
  Equations \textbf{23} (1998), no.~1{\&}2, 141--169.

\bibitem{DKriz1}
J.~Dittrich and J.~K{\v{r}}{\'{\i}}{\v{z}}, \emph{Bound states in straight
  quantum waveguides with combined boundary condition}, J. Math. Phys.
  \textbf{43} (2002), 3892--3915.

\bibitem{DKriz2}
\bysame, \emph{Curved planar quantum wires with {D}irichlet and {N}eumann
  boundary conditions}, J. Phys. A \textbf{35} (2002), L269--275.

\bibitem{DE}
P.~Duclos and P.~Exner, \emph{{C}urvature-induced bound states in quantum
  waveguides in two and three dimensions}, Rev. Math. Phys. \textbf{7} (1995),
  73--102.

\bibitem{DEK1}
P.~Duclos, P.~Exner, and D.~Krej\v{c}i\v{r}\'{\i}k, \emph{Locally curved
  quantum layers}, Ukrainian J.~Phys. \textbf{45} (2000), 595--601.

\bibitem{DEK2}
\bysame, \emph{Bound states in curved quantum layers}, Commun. Math. Phys.
  \textbf{223} (2001), 13--28.

\bibitem{Edmunds-Evans}
D.~E. Edmunds and W.~D. Evans, \emph{Spectral theory and differential
  operators}, Oxford University Press, New York, 1987.

\bibitem{MK-Kov}
T.~Ekholm and H.~Kova{\v{r}}{\'\i}k, \emph{Stability of the magnetic
  {S}chr{\"o}dinger operator in a waveguide}, math-ph/0404069.

\bibitem{ELV}
D.~V. Evans, M.~Levitin, and D.~Vassiliev, \emph{Existence theorems for trapped
  modes}, J.~Fluid Mech. \textbf{261} (1994), 21--31.

\bibitem{E4}
P.~Exner, \emph{Spectral properties of {S}chr{\"o}dinger operators with a
  strongly attractive {$\delta$} interaction supported by a surface},
  Contemporary Mathematics, AMS, vol. 339, Providence, R.I., 2003, Proceedings
  of the NSF Summer Research Conference (Mt. Holyoke 2002), pp.~25--36.

\bibitem{EFK}
P.~Exner, P.~Freitas, and D.~Krej\v{c}i\v{r}\'{\i}k, \emph{A lower bound to the
  spectral threshold in curved tubes}, R.~Soc. Lond. Proc. Ser.~A Math. Phys.
  Eng. Sci., to appear.

\bibitem{EGST}
P.~Exner, R.~Gawlista, P.~{\v S}eba, and M.~Tater, \emph{Point interactions in
  a strip}, Ann. Phys. \textbf{252} (1996), 133--179.

\bibitem{EI}
P.~Exner and T.~Ichinose, \emph{Geometrically induced spectrum in curved leaky
  wires}, J.~Phys A~ \textbf{34} (2001), 1439--1450.

\bibitem{EJKov1}
P.~Exner, A.~Joye, and H.~Kova\v{r}{\'\i}k, \emph{Edge currents in the absence
  of edges}, Phys. Lett. A~ \textbf{264} (1999), 124--130.

\bibitem{EJKov2}
\bysame, \emph{Magnetic transport in a straight parabolic channel}, J.~Phys. A~
  \textbf{34} (2001), 9733--9752.

\bibitem{ESylwia1}
P.~Exner and S.~Kondej, 
  \emph{Curvature-induced bound states for a {$\delta$} interaction
  supported by a curve in {$\mathbb{R}^3$}}, Ann. H.~Poincar{\'e} \textbf{3}
  (2002), 967--981.

\bibitem{ESylwia2}
\bysame, \emph{Bound states due to a strong {$\delta$} interaction supported by
  a curved surface}, J.~Phys A~ \textbf{36} (2003), 443--457.

\bibitem{ESylwia3}
\bysame, \emph{Strong-coupling asymptotic expansion for
  {S}chr{\"o}dinger operators with a singular interaction supported by a curve
  in {$\mathbb{R}^3$}}, 
  Rev.\ Math.\ Phys.~\textbf{16} (2004), no.~5, 559--582.

\bibitem{ESylwia4}
\bysame, 
  \emph{Schr\"odinger operators with singular interactions: 
  a model of tunneling resonances}, 
  J.~Phys.\ A~\textbf{37} (2004), 8255-8277.
 
\bibitem{ESylwia5}
\bysame, 
  \emph{Leaky quantum wire and dots: a resonance model}, 
  [math-ph/0307030].

\bibitem{ESylwia6}
\bysame, 
  \emph{Scattering by local deformations of a straight leaky wire}, 
  [math-ph/0410007].

\bibitem{EKov}
P.~Exner and H.~Kova\v{r}{\'\i}k, \emph{Magnetic strip waveguides}, J.~Phys. A~
  \textbf{33} (2000), 3297--3311.

\bibitem{EK1}
P.~Exner and D.~Krej\v{c}i\v{r}\'{\i}k, \emph{Quantum waveguides with a lateral
  semitransparent barrier: Spectral and scattering properties}, J.~Phys.~A
  \textbf{32} (1999), 4475--4494.

\bibitem{EK3}
\bysame, \emph{Bound states in mildly curved layers}, J.~Phys. A~ \textbf{34}
  (2001), 5969--5985.

\bibitem{EK2}
\bysame, \emph{Waveguides coupled through a semitransparent barrier: the
  weak-coupling behaviour}, Rev.~Math.~Phys. \textbf{13} (2001), no.~3,
  307--334.

\bibitem{EN}
P.~Exner and K.~N{\v{e}}mcov{\'a}, \emph{Quantum mechanics of layers with a
  finite number of point perturbations}, J.~Math. Phys. \textbf{43} (2002),
  1152--1184.

\bibitem{EN1}
\bysame, 
  \emph{Leaky quantum graphs: approximations by point interaction Hamiltonians}, 
  J.~Phys. A~\textbf{36} (2003), 10173-10193.
 
\bibitem{ES}
P.~Exner and P.~{\v S}eba, \emph{Bound states in curved quantum waveguides},
  J.~Math.~Phys. \textbf{30} (1989), 2574--2580.

\bibitem{ESS1}
P.~Exner, P.~{\v S}eba, and P.~{\v S}{\v{t}}ov{\'\i}{\v c}ek, \emph{On
  existence of a bound state in an {L}-shaped waveguide}, Czech. J. Phys. B~
  \textbf{39} (1989), 1181--1191.

\bibitem{ESTV}
P.~Exner, P.~{\v S}eba, M.~Tater, and D.~Van{\v e}k, \emph{{B}ound states and
  scattering in quantum waveguides coupled laterally through a boundary
  window}, J. Math. Phys. \textbf{37} (1996), 4867--4887.

\bibitem{EV}
P.~Exner and S.~A. Vugalter, \emph{Asymptotic estimates for bound states in
  quantum waveguides coupled laterally through a narrow window}, Ann. Inst. H.
  Poincar{\'e} \textbf{65} (1996), 109--123.

\bibitem{EV1}
\bysame, \emph{Bound-state asymptotic estimates for window-coupled dirichlet
  strips and layers}, J.~Phys. A~ \textbf{30} (1997), 7863--7878.

\bibitem{EV2}
\bysame, \emph{Bound states in a locally deformed waveguide: {T}he critical
  case}, Lett. Math. Phys. \textbf{39} (1997), 59--68.

\bibitem{EY1}
P.~Exner and K.~Yoshitomi, \emph{Band gap of the {S}chr{\"o}dinger operator
  with a strong {$\delta$}-interaction on a periodic curve}, Ann.
  H.~Poncar{\'e} \textbf{2} (2001), no.~6, 1139--1158.

\bibitem{FK}
P.~Freitas and D.~Krej\v{c}i\v{r}\'{\i}k,
\emph{A lower bound to the spectral threshold in curved strips
with Dirichlet and Robin boundary conditions},
submitted.

\bibitem{FTsutsui}
T.~F{\"u}l{\"o}p and I.~Tsutsui, \emph{A free particle on a circle with point
  interaction}, Phys. Lett.~A \textbf{264} (2000), no.~5, 366--374.

\bibitem{Glazman}
I.~M. Glazman, \emph{Direct methods of qualitative spectral analysis of
  singular differential operators}, Israel Program for Scientific Translations,
  Jerusalem, 1965.

\bibitem{GJ}
J.~Goldstone and R.~L. Jaffe, \emph{Bound states in twisting tubes}, Phys. Rev.
  B~ \textbf{45} (1992), 14100--14107.

\bibitem{Hurt}
N.~E. Hurt, \emph{Mathematical physics of quantum wires and devices}, Kluwer,
  Dordrecht, 2000.

\bibitem{Kl}
M.~Klaus, \emph{On the bound state of {S}chr{\" o}dinger operators in one
  dimension}, Ann. Phys. \textbf{108} (1977), 288--300.

\bibitem{KS}
F.~Kleespies and P.~Stollmann, \emph{Lifshitz asymptotics and localization for
  random quantum waveguides}, Rev. Math. Phys. \textbf{12} (2000), no.~10,
  1345--1365.

\bibitem{Kli}
W.~Klingenberg, \emph{A course in differential geometry}, Springer-Verlag, New
  York, 1978.

\bibitem{these}
D.~Krej\v{c}i\v{r}\'{\i}k, \emph{Guides d'ondes quantiques bidimensionnels},
  Ph.D. thesis, Facultas Mathematica Physicaque, Universitas Carolina
  Pragensis; Facult\'e des Sciences et Techniques, Universit\'e de Toulon et du
  Var, September 2001, Supervisors: P.~Duclos and P.~Exner.

\bibitem{K1}
\bysame, \emph{Quantum strips on surfaces}, J.~Geom. Phys. \textbf{45} (2003),
  no.~1--2, 203--217.

\bibitem{KT}
D.~Krej\v{c}i\v{r}\'{\i}k and R.~Tiedra de~Aldecoa, \emph{The nature of the
  essential spectrum in curved quantum waveguides}, J.~Phys.~A \textbf{37}
  (2004), no.~20, 5449--5466.

\bibitem{Kreyszig}
E.~Kreyszig, \emph{Differential geometry}, University of Toronto Press,
  Toronto, 1959.

\bibitem{Kriz}
J.~K{\v r}\'{\i}{\v z}, \emph{Spectral properties of planar quantum waveguides
  with combined boundary conditions}, Ph.D. thesis, Facultas Mathematica
  Physicaque, Universitas Carolina Pragensis, April 2003, Supervisor:
  J.~Dittrich.

\bibitem{KZ1}
P.~Kuchment and H.~Zeng, \emph{Convergence of spectra of mesoscopic systems
  collapsing onto a graph}, J.~Math.~Anal.~Appl. \textbf{258} (2001), 671--700.

\bibitem{KZ2}
\bysame, \emph{Asymptotics of spectra of {N}eumann {L}aplacians in thin
  domains}, Advances in differential equations and mathematical physics
  (Birmingham, AL, 2002), Contemp. Math., vol. 327, Amer. Math. Soc.,
  Providence, RI, 2003, pp.~199--213.

\bibitem{LL1}
Ch. Lin and Z.~Lu, \emph{Existence of bound states for layers built over
  hypersurfaces in~{$\mathbb{R}^{n+1}$}}, [math.DG/0402252].

\bibitem{LCM}
J.~T. Londergan, J.~P. Carini, and D.~P. Murdock, \emph{Binding and scattering
  in two-dimensional systems}, LNP, vol. m60, Springer, Berlin, 1999.

\bibitem{MFJ}
A.~Mekis, S.~Fan, and J.~D. Joannopoulos, \emph{Bound states in photonic
  crystal waveguides and waveguide bends}, Phys.\ Rev. B~ \textbf{58} (1998),
  4809--4817.

\bibitem{Newton}
R.~G. Newton, \emph{Bounds for the number of bound states for {S}chr{\"o}dinger
  equation in one and two dimensions}, J.~Operator Theory \textbf{10} (1983),
  119--125.

\bibitem{OM}
O.~Olendski and L.~Mikhailovska, \emph{Localized-mode evolution in a curved
  planar waveguide with combined {D}irichlet and {N}eumann boundary
  conditions}, Phys. Rev. E~ \textbf{67} (2003), art.~056625.

\bibitem{RS1}
M.~Reed and B.~Simon, \emph{Methods of modern mathematical physics,
  {I}.~{F}unctional analysis}, Academic Press, New York, 1972.

\bibitem{RS4}
M.~Reed and B.~Simon, \emph{Methods of modern mathematical physics,
  {IV}.~{A}nalysis of operators}, Academic Press, New York, 1978.

\bibitem{Rellich}
F.~Rellich, \emph{Das {E}igenwertproblem von {$\Delta u+\lambda u=0$} in
  {H}albr{\"{o}}hren}, Studies and Essays Presented to {R}.~{C}ourant on his
  60th Birthday, {J}anuary 8, 1948, Interscience Publishers, Inc., New York,
  1948, pp.~329--344.

\bibitem{RB}
W.~Renger and W.~Bulla, \emph{Existence of bound states in quantum waveguides
  under weak conditions}, Lett.~Math.~Phys. \textbf{35} (1995), 1--12.

\bibitem{Seto}
N.~Seto, \emph{Bargmann's inequalities in spaces of arbitrary dimension}, Publ.
  RIMS \textbf{9} (1974), 429--461.

\bibitem{SS}
E.~Shargorodsky and A.~V. Sobolev, \emph{Quasi-conformal mappings and periodic
  spectral problems in dimension two}, J.~Anal. Math. \textbf{91} (2003),
  67--103.

\bibitem{Weidmann}
J.~Weidmann, \emph{Linear operators in {Hilbert} spaces}, Springer-Verlag, New
  York Inc., 1980.

\bibitem{Y}
K.~Yoshitomi, \emph{Band gap of the spectrum in periodically curved quantum
  waveguides}, J.~Differential Equations \textbf{142} (1998), no.~1, 123--166.

\end{thebibliography}
%
\providecommand{\bysame}{\leavevmode\hbox to3em{\hrulefill}\thinspace}
\providecommand{\MR}{\relax\ifhmode\unskip\space\fi MR }
\providecommand{\MRhref}[2]{%
  \href{http://www.ams.org/mathscinet-getitem?mr=#1}{#2}
}
\providecommand{\href}[2]{#2}

%
%
%
\end{document}